\documentclass[final,5p,times,twocolumn,authoryear]{elsarticle}
\usepackage{xcolor}
\usepackage{graphicx} 
\usepackage{float}
\usepackage{amsmath, amssymb, mathrsfs}
\usepackage{lipsum}
\usepackage{url}
\usepackage[english]{babel}
\usepackage{nameref}
\usepackage{hyperref}

\journal{High Energy Astrophysics}

\begin{document}

\begin{frontmatter}

\title{Probing cosmic isotropy: Hubble constant and matter density large-angle variations with the Pantheon+SH0ES data}

\author[first]{Rahima Mokeddem } 
\ead{rahima.mookeddem@inpe.br}

\affiliation[first]{Instituto Nacional de Pesquisas Espaciais, Divisao de Astrofísica
,
            addressline={Av. dos Astronautas, 1758}, 
            city={São José dos Campos},
            postcode={12227-010}, 
            state={SP},
            country={Brazil}}
            
\author[second]{Maria Lopes, Felipe Avila, Armando Bernui}
\ead{marialopes@on.br, fsavila2@gmail.com, bernui@on.br}
\affiliation[second]{Observatório Nacional,
            addressline={Rua General José Cristino, 77, São Cristóvão}, 
            city={Rio de Janeiro},
            postcode={20921-400}, 
            state={RJ},
            country={Brazil}}

 \author[fourth,fiveth]{Wiliam S. Hipólito-Ricaldi}
 \ead{wiliam.ricaldi@ufes.br}
\affiliation[fourth]{Departamento de Ciencias Naturais, Universidade Federal do Espírito Santo,
            addressline={Rodovia BR 101 Norte, km. 60}, 
            city={São Mateus},
            postcode={29932-540}, 
            state={ES},
            country={Brazil}}
\affiliation[fiveth]{Nucleo Cosmo-UFES, Universidade Federal do Espírito Santo,
            addressline={Av. Fernando Ferrari, 540}, 
            city={Vitória},
            postcode={29075-910}, 
            state={ES},
            country={Brazil}}            

\begin{abstract}
In this study we investigate potential large-angle anisotropies in the angular distribution of the 
cosmological parameters $H_0$ (the Hubble constant) and $\Omega_m$ (the matter density) in the flat-$\Lambda$CDM framework, using the Pantheon+SH0ES supernovae  catalog. 
For this we perform a directional analysis by dividing the celestial sphere into a set of directions, and estimate the best-fit cosmological parameters across the sky using a MCMC approach. 
Our results show a dominant dipolar pattern for both parameters in study, suggesting a preferred axis in the universe expansion and in the distribution of matter. 
However, we also found that for $z \gtrsim 0.015$, this dipolar behavior is not statistically significant, 
confirming the expectation --in the $\Lambda$CDM scenario-- of an isotropic expansion and a uniform 
angular distribution of matter 
(both results at $1\,\sigma$ confidence level). 
Nevertheless, for nearby supernovae, at distances $\lesssim 60$ Mpc or $z \lesssim 0.015$, 
the peculiar velocities introduce a highly significant dipole in the angular distribution of $H_0$. 
Furthermore, we perform various robustness tests that support our findings, and consistency tests of our methodology.
\end{abstract}


\end{frontmatter}

\section{Introduction}~\label{sec:intro}
In the precision cosmology era, where accurate datasets from large and deep astronomical surveys are available, the task of finding the cosmological model that reproduces all these data is indeed challenging~\citep{perivolaropoulos2022challenges, riessetal, divalentino, Fosalba_2021}. 

Currently, the flat-$\Lambda$CDM model seems to fulfill this objective, despite the fact that according to it, 95\% of the components of the universe remain unknown~\citep{Planck2020-VI, peebles, Frieman_2008}. 
In the absence of strong competitors, $\Lambda$CDM stays as the concordance cosmological model, although our ignorance regarding the physical nature of the dark sector is an uncomfortable situation~\citep{BULL2016,Verde_2019,Di_Valentino_2021,Eleonora2025, Luongo_2022}.

One  interesting remaining task is to test the model consistency by comparing the expected $\Lambda$CDM properties, with the corresponding observed phenomenon using updated cosmological data~\citep{Linder2021}. In that sense, the analyses performed in this work allow us to: 
i)~confirm that the flat-$\Lambda$CDM remains the concordance model given the current precision of the data regarding its (statistically) homogeneous and isotropic properties; 
ii)~assess the statistical probability that current precision cosmology has outgrown the $\Lambda$CDM paradigm (see, e.g.,~\cite{Aluri2023, Krishnan_2021}); 
iii)~identify observational systematics that may be impacting the analyses.

Perhaps the main feature of the flat-$\Lambda$CDM model is the competence to determine $\Omega_m$\footnote{Throughout the work, we will adopt the following definition: $\Omega_m(z=0)\equiv\Omega_{m0}\equiv \Omega_m$.}, the parameter that measures the amount of dark plus baryonic matter observed today. 
One natural expectation regarding $\Omega_m$ is that measurements along different sky directions performed with the same cosmic tracer should roughly show equal values of $\Omega_m$, except for fluctuations due to measurement uncertainties, a consequence of the expected isotropic matter distribution~\citep{Javanmardi2015}. 
A similar analysis can be performed considering the Hubble constant $H_0$, the expansion rate measured today, another important parameter of the concordance cosmological model. 
Recent studies have suggested potential anisotropies in $H_0$, including a quadrupolar pattern found in Pantheon+ data at the $2\sigma$ level~\citep{Cowell2023}, indicating deviations from the expected statistical isotropy
In fact,~\cite{hu2024a,hu2024b} analyzed cosmic anisotropies using Pantheon+SH0ES sample, reporting variations in the Hubble constant and other parameters. 

In addition,~\cite{perivolaropoulos2022challenges} investigated modern challenges to the flat-$\Lambda$CDM model, focusing on  possible large-scale anisotropy in cosmological parameters, including $H_0$ and $\Omega_m$. 
However, significant deviations of $\Omega_m$ or $H_0$ in different directions are not expected because the flat-$\Lambda$CDM model is based on the cosmological 
principle~\citep{Maartens2011,Appleby2014,
Schwarz2015,Avila19,Dias23,Kester24,Franco24,Franco25}.

Recent literature reports interesting results regarding this problem, employing diverse methodologies to investigate several datasets. For example, 
\cite{conville2023} using hemispheres to scan the sky found angular variations, up to $4$ km/s/Mpc, in the Hubble constant $H_0$. 
Their analyses, at different redshift intervals, intended to measure the Hubble constant absolute difference, defined as $\Delta H_0 \equiv H_{0}^{N}-H_{0}^{S}$, where $N$ and $S$ means North and South hemispheres, respectively. 
In addition to SNe Ia analyses, several independent observational probes have also reported possible anisotropies in the local expansion rate. Using galaxy clusters at distances below $\sim 500$ Mpc, the works of
\cite{Migkas_2020, Migkas2021} found statistically significant directional variations in the X-ray luminosity temperature relation, which can be interpreted as variations in $H_0$. Owing to their large cluster statistics and extensive sky coverage, these studies provide complementary evidence for deviations from perfect isotropy on intermediate scales.
Furthermore, the recent Cosmicflows-4 (CF4) catalog and its upgrades have revealed additional indications of large-scale anisotropy. The dynamical scale of homogeneity appears not to be reached by $\sim 300\,h^{-1}\,\mathrm{Mpc}$ \citep{Courtois2025}, and analyses of the same dataset have reported nearly $4\sigma$ variations in the inferred Hubble constant \citep{Boubel_2025}. These findings are consistent 
with earlier detections of an anomalous bulk flow within CF4 \citep{Watkins2023}, suggesting that coherent peculiar 
motions may play a non-negligible role in shaping local measurements of cosmological parameters. Even within the SN Ia literature, directional variations in $H_0$ have been 
discussed previously. Notably, \cite{Krishnan_2022, Zhai_2022}
found hemispherical differences in $H_0$, with the latter  incorporating Cepheid-calibrated supernovae and thus probing the lowest rung of the distance ladder. These studies reinforce the importance of exploring the angular 
dependence of cosmological parameters using the updated Pantheon+SH0ES data, which offers improved calibration and a unified covariance treatment.

Other studies of SNIa, performed directional analyses in thin redshift bins finding that overdense and underdense structures in the Local Universe cause deviations from the expected statistical isotropy (see, e.g.,~\cite{perivolaropoulos2023,tang2023,Lopes2024,Sah24}). Studies of the angular distribution of the cosmological 
parameters $H_0$ and $\Omega_m$, using the Pantheon+SH0ES dataset, were done in an effort to 
map a local matter underdensity region responsible for a preferred direction of 
cosmic anisotropy~\citep{hu2024a,hu2024b}.
Assuming the flat-$\Lambda$CDM model,~\cite{clocchiatti2024} carried out an angular analysis focusing on how the $\Omega_{\Lambda}$ parameter varies with direction. They found an anisotropy that is interpreted as an apparent effect associated with the relativistic frame of reference transformation~\citep{Tsagas11}. 

Complementing these works, recent full-sky analysis of the Pantheon+ SNe sample done in several redshift bins reported that the $\Lambda$CDM parameters $(H_0,\Omega_m)$ evolve with redshift~\citep{Malekjani2024}, suggesting statistical fluctuation, or unexplored systematics in the data, or   a breakdown of the $\Lambda$CDM model.

In this work we use the Pantheon+SH0ES catalog to study possible large-angle anisotropies associated to deviations of the cosmological parameters $H_0$ and $\Omega_{m}$ with respect to the expected values in the flat-$\Lambda$CDM model. 
Throughout this study, we adopted an approach based on the analysis of the distance modulus of SNe located within hemispheres. 
This allows us to perform a directional analysis over the sky, searching for directions where anomalous $\Omega_m$ and $H_0$ deviations could manifest. 
Our results show a dipolar pattern for the cosmological parameters in study, i.e., $H_0$ and $\Omega_m$, suggesting a preferred axis in the universe expansion and in the distribution of matter. 
For this reason, we also investigate if this dipolar behavior is consistent with what is expected in the flat-$\Lambda$CDM model. 
In fact, the statistical significance of large-angle anisotropies will be quantified by comparison with a large set of simulated isotropized maps, as described below. 

This work is structured as follows: Section~\ref{data} introduces the Pantheon+SH0ES catalog and outlines its key properties relevant to our analyses. In Section~\ref{methodology}, we detail the methodology employed to select the Type Ia supernovae (SNe) for determining $H_0$ and $\Omega_m$, as well as the construction of isotropic maps and the covariance matrix, which are critical components of our analysis. 
Our findings are presented in Section~\ref{results}, followed by a discussion of the conclusions and final remarks in Section~\ref{Summary and conclusions}. 
Additionally, all robustness tests supporting our main results are provided in the Appendices. 


\section{{Observational Data: The Pantheon+SH0ES Catalog}} \label{data}

Supernovae (SNe) events are transient and appear randomly on the sky. 
Because type Ia supernovae (SNe Ia) are standardizable candles, efforts to calibrate their light-curves results in high quality compilations that are publicly available. 
In this study we used the Pantheon$+$ SH0ES catalog~\footnote{\url{https://github.com/PantheonPlusSH0ES/DataRelease}} \citep{brout2022,Scolnic2022}, the successor to the original Pantheon compilation of SNe Ia events \citep{Scolnic2018}. 
The Pantheon+SH0ES catalog contains 1550 SNe Ia events and 
1701 supernovae light-curves; 
it includes those SNe located in neighboring host galaxies whose distances have been determined using Cepheids. 
From now on, we will refer to supernovae light-curves simply as SNe. 

\begin{figure}[htbp]
\centering
\includegraphics[width=\linewidth]{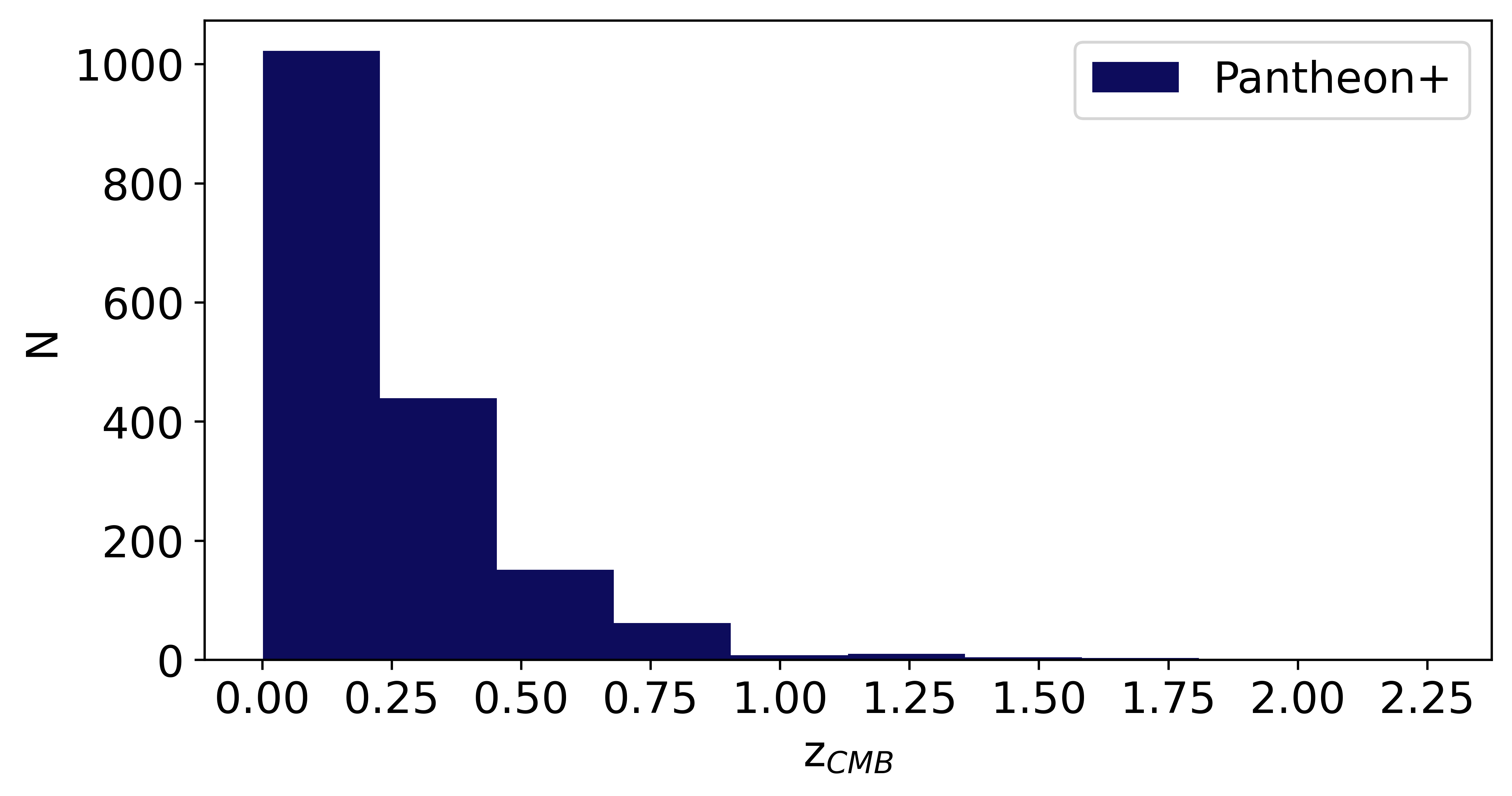}
\caption{Histogram of the redshift distribution of the Pantheon+SH0ES supernovae catalog, in the CMB redshift frame.}
\label{fig:histogram_SNe}
\end{figure}

\begin{figure}[htbp]
    \centering
    \includegraphics[width=\linewidth]{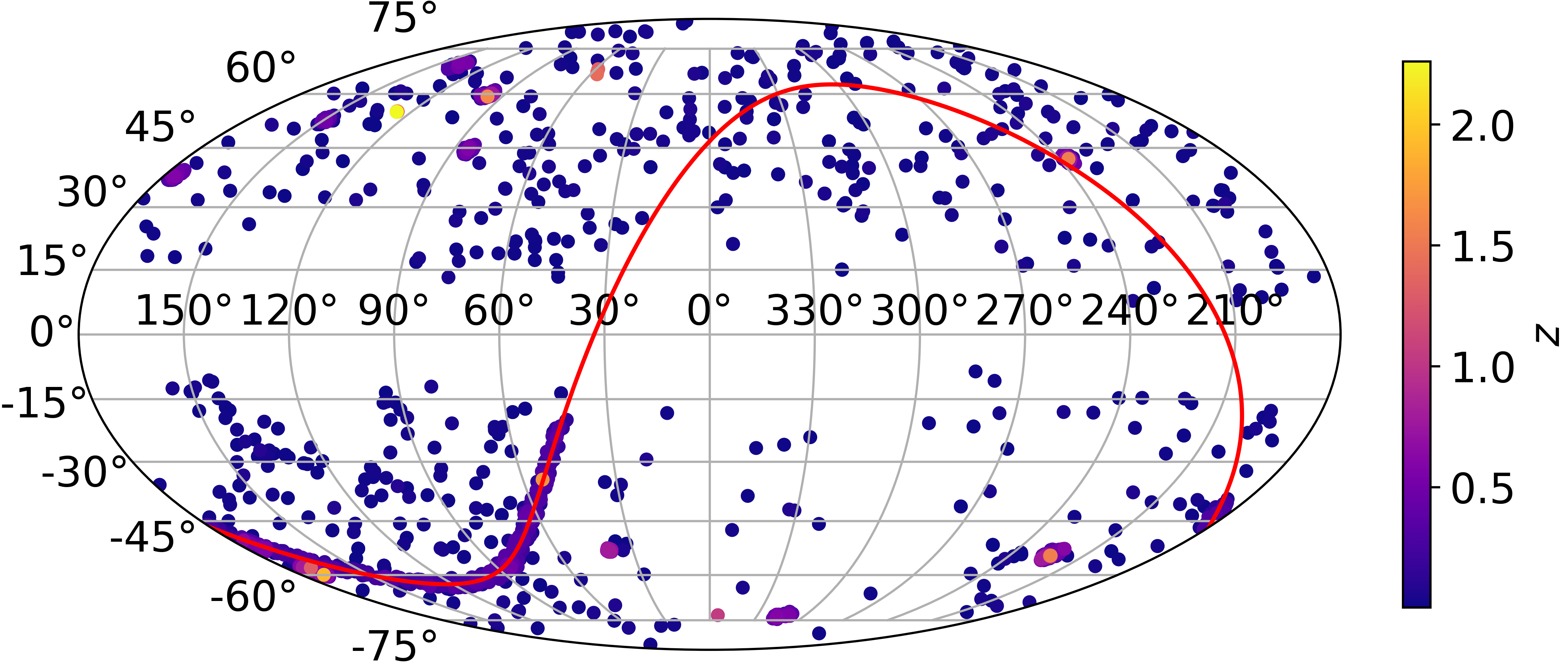}
\caption{
Mollweide projection in galactic coordinates of the Pantheon+SH0ES supernovae events, represented as colored dots, on the celestial sphere. 
The colors of the SNe events represent the redshift of the galaxy host. 
The solid red line represents the celestial equator.
}
\label{fig:footprint_SNe}
\end{figure}

The redshift range of these SNe is $0.001 \leq z_{\,\text{CMB}} \leq 2.261$ 
(where CMB stands for the Cosmic Microwave Background frame of reference; 
in what follows we use $z \equiv z_{\,\text{CMB}}$), 
with distribution shown in Figure~\ref{fig:histogram_SNe} 
and its sky footprint is displayed in 
Figure~\ref{fig:footprint_SNe}. 
The comprehensive collection of precise data obtained through spectroscopy includes the distance covariance matrix~\citep{Scolnic2022}, which contains all the correlations from SNe duplications and the distance measurements 
due to several systematic uncertainties.

To investigate the large-scale features in the parameters $\Omega_{m}$ and $H_0$, we consider three samples with redshift intervals 
$[z_{\text{min}},z_{\text{max}}]$, where 
$z_{\text{max}}=2.261$ is fixed, and $z_{\text{min}}$ takes the values $0.01$, $0.015$, and $0.02$. 
The number of SNe in each redshift interval is: 
$1588, 1524$, and $1426$, respectively.

In what follows we present the results for the case $z_{\text{min}} = 0.015$, leaving the other two cases for the Appendix section. 

The methodology applied in our study uses the following SNe data to obtain the cosmological parameters: 
the SN sky position (\textbf{RA}, \textbf{DEC}), the CMB redshift \textbf{zCMB}, the standardized distance modulus \textbf{MU\_SH0ES}, the standardized \textbf{m\_b} magnitude \textbf{m\_b\_corr}, 
and the distance covariance matrix \href{https://github.com/PantheonPlusSH0ES/DataRelease/blob/main/Pantheon%2B_Data/4_DISTANCES_AND_COVAR/Pantheon%2BSH0ES_STAT%2BSYS.cov}{\textbf{Pantheon+SH0ES\_STAT\_SYS.cov}}.
As suggested in the literature, we used the CMB reference frame because it is the suitable frame to investigate the large-angle variations of the cosmological parameters in study.


\section{Methodology} \label{methodology}

In this section we describe the methodology used for the directional analyses to estimate the large-angle variations of the cosmological parameters $H_0$ and $\Omega_m$ within the flat-$\Lambda$CDM model. 
Our methodology consists of three main steps. 
Firstly, we divide the sky into a set of hemispheres covering the entire celestial sphere. 
Then, we perform Monte Carlo Markov Chain (MCMC) analyses using the MCMC ensemble sampler \textbf{emcee}~\citep{Foreman-Mackey2013, Foreman-Mackey2019} in each hemisphere to estimate $H_0$ and $\Omega_m$. 
Finally, the statistical significance of our results is evaluated by comparing them with the results obtained from similar analyses applied to a large set of simulated data, in which the SNe distance modulus data were randomized. 
Below, we present a detailed discussion of each one of these steps.

\subsection{Directional analysis of SNe data}
\label{directional}

Consider the $J$-th spherical cap, 
$\mathscr{C}^J_{\gamma}$, with center in $(\theta_J,\phi_J)$ and radius $\gamma$, and define a scalar function to associate a non negative real value to the corresponding cap, that is 
\begin{equation}
\text{P}^{\,J}: \mathscr{C}^J_{\gamma} \subset \mathcal{S}^2 \mapsto \mathbb{R}^+ \,,
\end{equation}
for $J = 1, \cdots, N_{\textrm{\small caps}}$, where $\text{P}^{\,J}$ denotes one cosmological parameter in study, $H_0^{J}$ or $\Omega_m^{J}$, obtained from the statistical analysis of the SNe (described below) located in the $J$-th spherical cap $\mathscr{C}^J_{\gamma}$. 
We shall adopt hemispheres, i.e, $\gamma = 90^{\circ}$, and denote the
$J$-th hemisphere just as $\mathscr{C}^J$. 
The set of $N_{\text{caps}}$ values, $\{ \text{P}^{\,J} \}$ for 
$J = 1, \cdots, N_{\textrm{\small caps}}$, 
is then assembled together into two full-sky maps, hereafter the $H_0$-map and the $\Omega_m$-map, for the corresponding parameter in analysis. 
Let $n_J$ be the number of SNe, listed in the Pantheon+SH0ES catalog, present in the $J$-th hemisphere. 
In this work, we used $N_{\text{caps}}=48$ and thus our analyses were performed in 48 hemispheres. 
As a robustness test, in the Appendix \ref{app:high_reso} we show the case with $N_{\text{caps}}=192$, where similar results were obtained. 


\subsection{\texorpdfstring{$H_0$}{} and \texorpdfstring{$\Omega_{\text{m}}$}{} using MCMC}

To estimate the $H_0^{J}$ or $\Omega_m^{J}$  parameters in each hemisphere, we use the MCMC method. The MCMC is a stochastic algorithm designed to sample from the posterior distribution, which is proportional to the product of the likelihood function and the prior probabilities (see e.g. ~\citet{PhysRevD.66.103511,Gelman2013}) 
\begin{equation}
\mathcal{P}(\theta|D) \propto \mathcal{L}(D|\theta) \mathcal{P}(\theta) \,,
\end{equation}
where $D$ is the dataset and  $\theta$ is the set of parameters. The MCMC method is appropriately applied  to the data collected in each of 
the 48 hemispheres. 
For each hemisphere,  we select the SNe Ia present and analyze their observed apparent magnitudes, $m_b$, which are related to the cosmological distance 
modulus $\mu$ by $m_b = \mu + M_B$, where $M_B$ is the absolute magnitude of SNe. 
We then fit these observational data points using  the cosmological distance modulus, which is a function of the luminosity distance $d_L$~\citep{Riess_1998}, 
\begin{equation}\label{eq-mu}
\mu(z)  = 5\,log_{10}\left(\frac{d_L(z)}{\text{Mpc}}\right) + 25 \,,
\end{equation}
where $d_L$ depends on the assumed cosmological model  
\begin{equation}
d_L(z) = c\,(1+z)\int_0 ^z \frac{dz'}{H(z')} \,,
\end{equation}
where 
\begin{equation}\label{H(z)}
H(z) = H_0 \sqrt{\Omega_m(1+z)^3 + \Omega_{\Lambda}} \,,
\end{equation}
with $H_0$ denoting the Hubble constant. 
Clearly, for the flat-$\Lambda$CDM model one has $\Omega_{\Lambda} = 1 - \Omega_m$, 
where $\Omega_m$ and $\Omega_{\Lambda}$ are the matter density and dark energy density parameters, respectively. The observed apparent magnitudes, $m^{\text{obs}}_{b}$, are compared to their theoretical counterparts, $m^{\text{theo}}_{b}$, derived from the $\Lambda$CDM model. The likelihood function used is defined as 
\begin{equation}\label{eq:likelihood}
 \mathcal{L}(D|\theta) \propto \exp\left[-\frac{1}{2}\chi^{2}(D|\theta)\right], 
\end{equation}
and the $\chi^{2}$  is given by
\begin{eqnarray}\label{chi2}
\chi^2(D|\theta) &=& \sum_{i,j} \left[m^\text{obs}_{b}(z_i) - m^\text{theo}_{b}(z_i; \theta)\right] \times C_{ij}^{-1} \nonumber \\
&\times& \left[m^\text{obs}_{b}(z_j) - m^\text{theo}_{b}(z_j; \theta)\right] \,,
\end{eqnarray}
where $\theta=\{H_0, \Omega_m, M_b\}$, and $C$ denotes the covariance matrix.


In our all-sky analysis we used the full \textit{Pantheon+SH0ES} covariance matrix $C$ with dimension $N \times N$ (the Pantheon+Shoes catalog contains 
$N=1701$ light-curve SNe analyses). 
This matrix incorporates both statistical and systematic uncertainties for 
all SNe, including the correlations introduced by calibration and by 
the Cepheid-host SNe used in the SH0ES distance-ladder 
calibration~\citep{brout2022}. 
This unified statistical + systematics covariance data guarantees that cosmological fits consistently propagate all measurement errors and their correlations, thereby enabling a direct and coherent joint estimation of 
$H_0$ and other cosmological parameters.

For each subset of SNe, defined by hemispheres or redshift bins, we consider a subset  $h \subset \{1,2,3\dots N\}$. 
The corresponding submatrix $C_h$ of the full covariance matrix $C$ is then 
obtained by 
\begin{eqnarray}
(C_h)_{ij}=C_{i_hj_h}\,, \qquad i_h,j_h \in h
\end{eqnarray}
where $i_h$ and $j_h$ denote the indices of the supernovae in the  subset $h$. This construction ensures that both diagonal elements (individual statistical uncertainties) and off-diagonal elements (correlated systematic contributions) are consistently preserved. 
The restricted cosmological likelihood for each subset is then evaluated by assuming a multivariate Gaussian distribution, guaranteeing that parameter estimation properly accounts for both statistical noise and correlated systematics~\citep{Scolnic2022}.

It is worth mentioning that the dependency between $H_0$ and  $M_B$ requires the inclusion of $M_B$ in the set of parameters for proper quantification (see, e.g.,~\cite{Staikova2023}). 
In the Pantheon+SHOES dataset, $M_B$ is constrained by 77 SNe in Cepheid-host galaxies, which are included in our hemispherical divisions according to their sky positions.Thus, $M_B$  is included as a third parameter in our analysis, but
we do not assemble a sky map with the set of values 
$\{ M^J_B \}$.

Once the MCMC for each of the 48 hemispheres is complete, the resultant set of values $\{ H^J_0 \}$ and $\{ \Omega^J_m \}$ 
in the sky directions are used to assemble the respective sky maps, namely the $H_0$- and the $\Omega_m$-maps 
(see Sec. \ref{directional}).

After that, the directional features of these maps can be analyzed in the harmonic space representation. 
Indeed, our analysis of the angular power spectrum 
of these maps quantifies the angular distribution of the 
$H_0$ and $\Omega_m$ parameters, revealing, in particular, potential dipolar anisotropies, suggestive of preferred directions.


\subsection{Simulating Isotropic \texorpdfstring{$H_0$}{} and \texorpdfstring{$\Omega_{\text{m}}$}{} maps} \label{isotropic-maps}

The final, and equally important, step of our approach is to evaluate the statistical significance of the multipole features of the $H_0$- and $\Omega_m$-maps, 
allowing us to assess the presence of anomalous deviations from statistical isotropy. 
As mentioned above, we focused our analyses on the sample with 
$z_{\text{min}} = 0.015$, which consists of $1524$ SNe. The evaluation  is carried out by comparing the corresponding angular power spectra of the data maps, namely the $H_0$- and $\Omega_m$-maps, with the angular power spectra computed from 
two sets of $1000$ simulated maps, $\{ H_0^{\text{ISO}} \}$-maps and the $\{ \Omega_m^{\text{ISO}} \}$-maps, 
produced following the isotropization procedure described below. This comparison allows us to evaluate  the statistical significance of the angular characteristics of the data maps.

To produce each simulated map, we first generate an isotropized distance 
modulus dataset. 
This is obtained in a two steps procedure applied to the Pantheon+SH0ES 
distance modulus dataset, 
$\{ \mu_i(z) \}, i=1,\cdots,1524$, that is, 
\begin{equation}
\mu_i \,\stackrel{\text{rand.}}{\rightarrow}\, 
\mu_i^{\text{ran}} \,\stackrel{\text{Gaus.}}{\rightarrow}\, \mu_i^{\text{ran+Gau}}\,, 
\end{equation}
preserving the number of SNe in each hemisphere, $\{ n_J \}$, 
for $J=1,\cdots,48$. 
%

The randomized set $\{ \mu_i^{\text{ran}} \}$ is produced by randomizing the distance moduli data of the SNe sample, $\{ \mu_i \}$, but maintaining their angular positions --and any other information regarding them-- fixed. 
In this way, this procedure eliminates any correlation between the sky direction of the SN event and its original distance modulus, maintaining the original angular distribution of the SNe sample. 

Then, each of these values is modified by adding a random value drawn from a Gaussian distribution with mean $\mu_i^{\text{ran}}$ and standard deviation equal to its measured uncertainty $\sigma_{\mu_i}$, resulting in the isotropized dataset $\{ \mu_i^{\text{ran+Gau}} \}$. 
The realization providing the set of triplets $\{ (\alpha_i,\delta_i,\mu_i^{\text{ran+Gau}}) \}, \,i=1,\cdots,1524$, form one simulated catalog. After applying our directional analysis and 
$\chi^2$ best-fitting procedures, this catalog produces a pair of maps: 
$H_0^{\text{ISO}}$-map and $\Omega_m^{\text{ISO}}$-map. 
To efficiently fit these cosmological parameters in each hemisphere, we employed the \textbf{Core Cosmology Library (CCL)}\footnote{\url{https://github.com/LSSTDESC/CCL}}\citep{Chisari2019}, assuming  a flat-$\Lambda$CDM cosmology\footnote{The fit for the distance modulus use the following Planck 2018 parameters \citep{Planck2020-VI}: $\Omega_b = 0.0494$ (baryonic matter density fraction), $\sigma_8 = 0.8120$ (matter density perturbation variance at 8 Mpc/$h$ scale), and $n_s = 0.9649$ (scalar spectral index).}. 
Finally, we repeat this procedure to obtain two sets of $1000$ maps each: the $\{ H_0^{\text{ISO}} \}$-maps and the 
$\{ \Omega_m^{\text{ISO}} \}$-maps, from which the angular power spectra of the ensembles $\{ H_0^{\text{ISO}} \}$ and 
$\{ \Omega_m^{\text{ISO}} \}$ can be computed. 


\section{Results and Discussions} \label{results}

We applied the methodology outlined in Sec.~\ref{methodology} to the subsets described in Sec.~\ref{data}. 
The results of these analyses,  for 48 hemispheres and $z_{min}= 0.015$, are presented in this section. 
Nevertheless, in the ~\ref{app:high_reso} we present the analysis for 192 hemispheres, and also considering other $z_{min}$ cases to evaluate the robustness of our findings. 

\begin{figure}[htbp]
\centering
\centering
\includegraphics[width=0.78\linewidth]{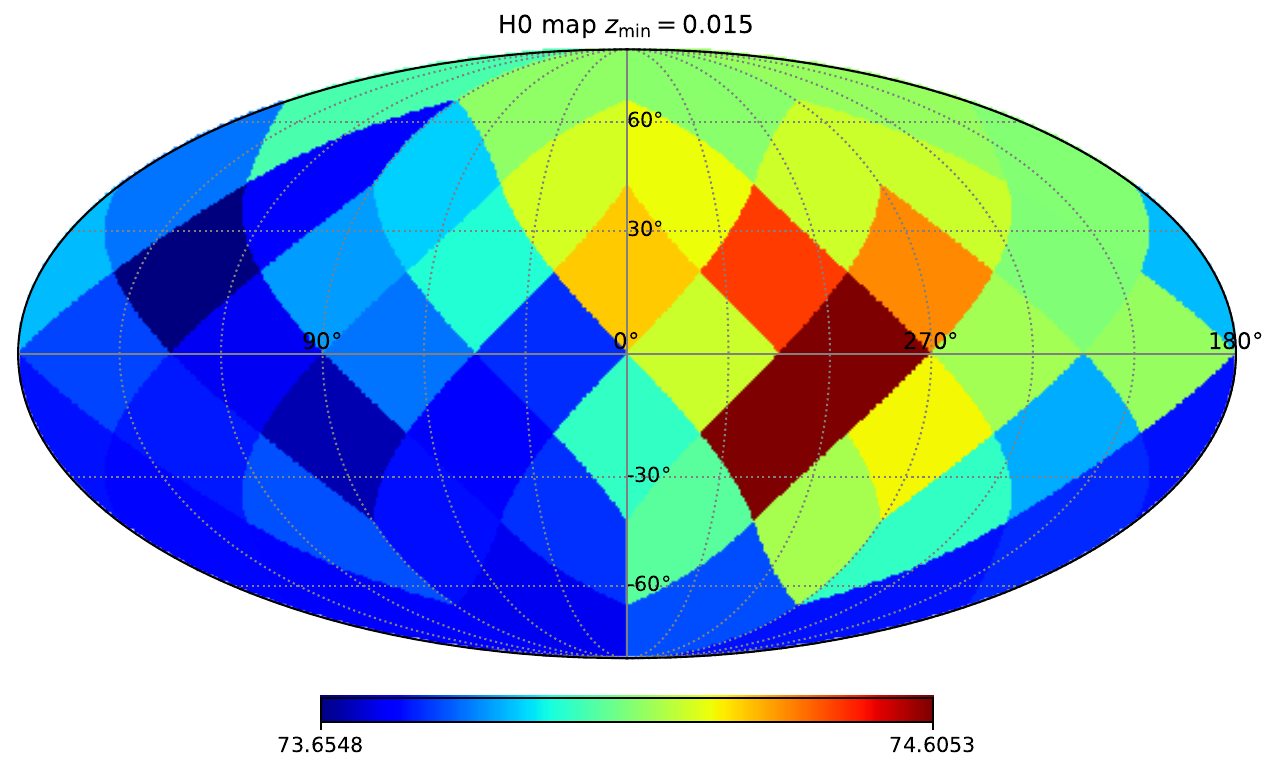}
\includegraphics[width=0.78\linewidth]{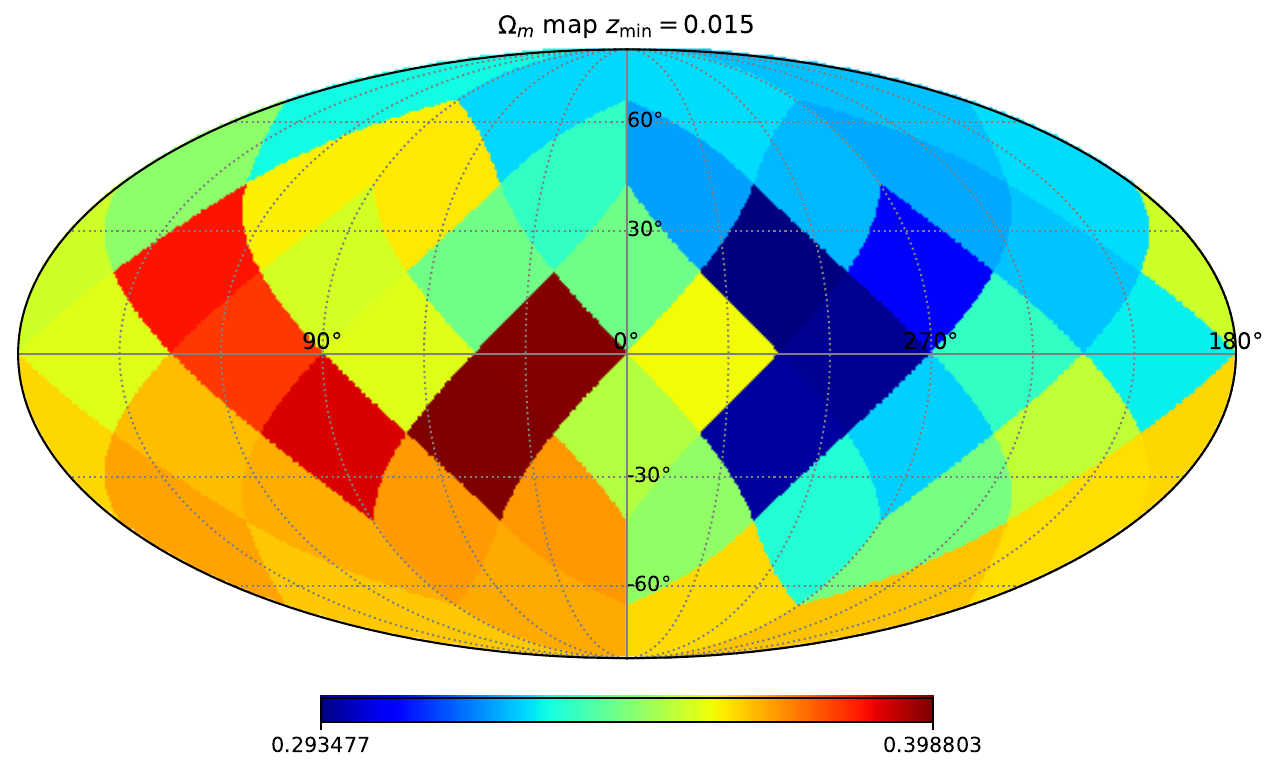}

\caption{
The $H_0$-map (upper map) and the $\Omega_m$-map (lower map), shown in Galactic coordinates,  resulting from our directional analysis of the redshift bin with $z_{\text{min}} = 0.015$,  considering 48 hemispheres. Grid lines mark Galactic longitudes and latitudes in $30^{\circ}$ intervals.
}
\label{fig:h0_map_0p015}
\end{figure}

\begin{figure}[htbp]
\centering
\includegraphics[width=0.78\linewidth]{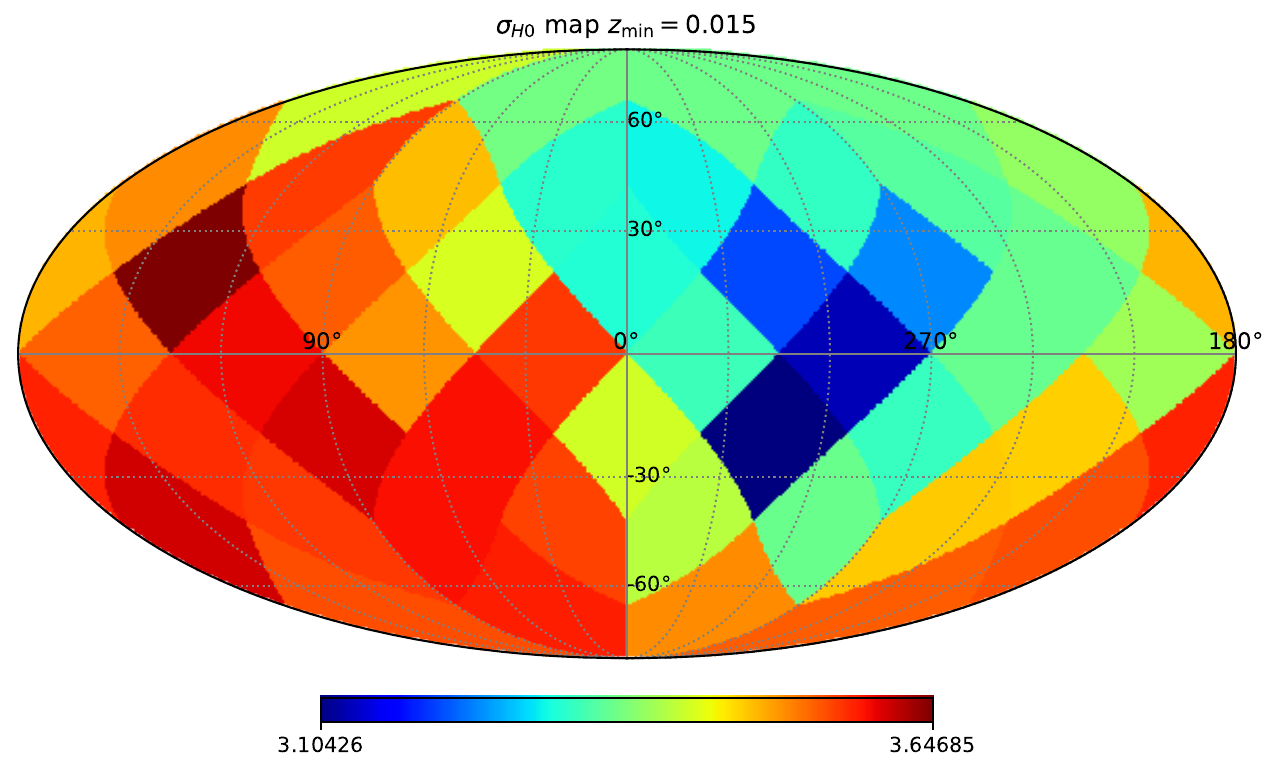}
\includegraphics[width=0.78\linewidth]{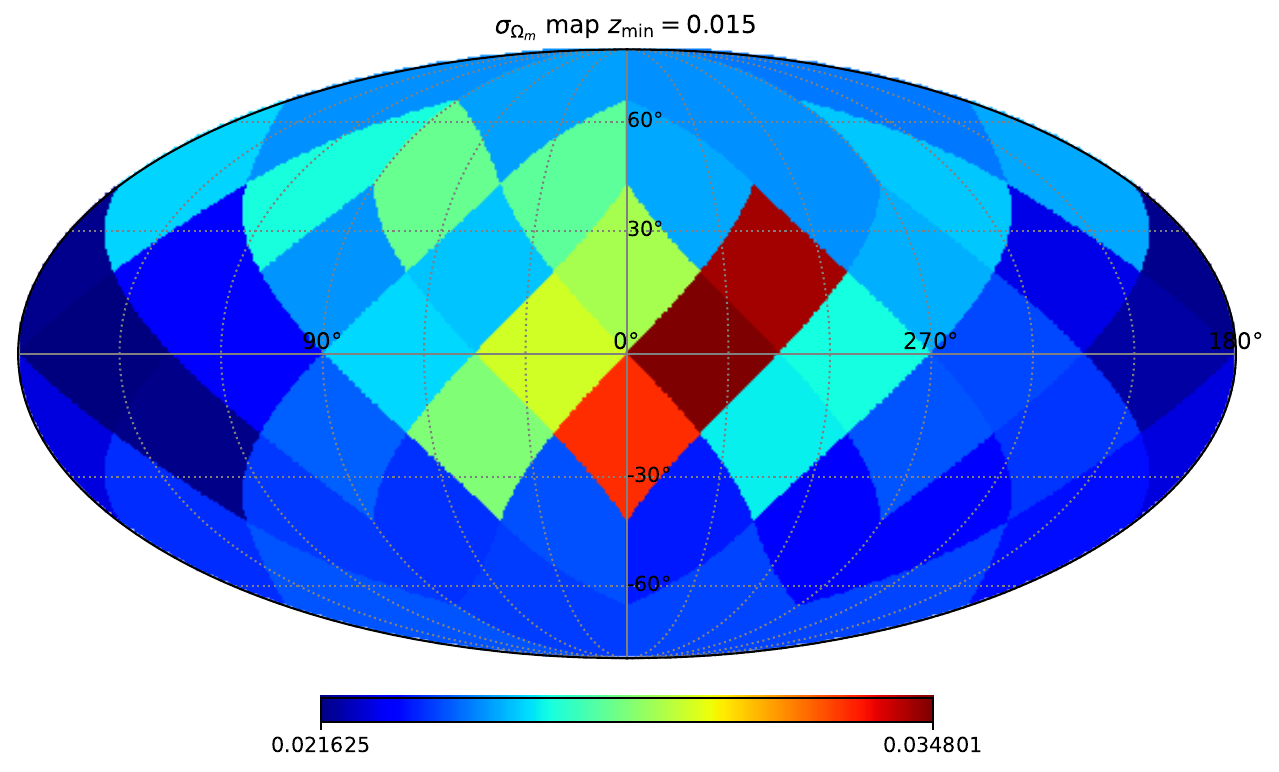}
\caption{The standard deviation maps: 
$\sigma_{H_0}$-map and $\sigma_{\Omega_m}$-map in Galactic coordinates. 
These analyses resulted from the study of the bin redshift with $z_{\text{min}} = 0.015$, 
and considering 48 hemispheres.
}
\label{fig:sigmas_map_48}
\end{figure}

\begin{figure}[htbp]
\centering  
\includegraphics[width=\columnwidth]{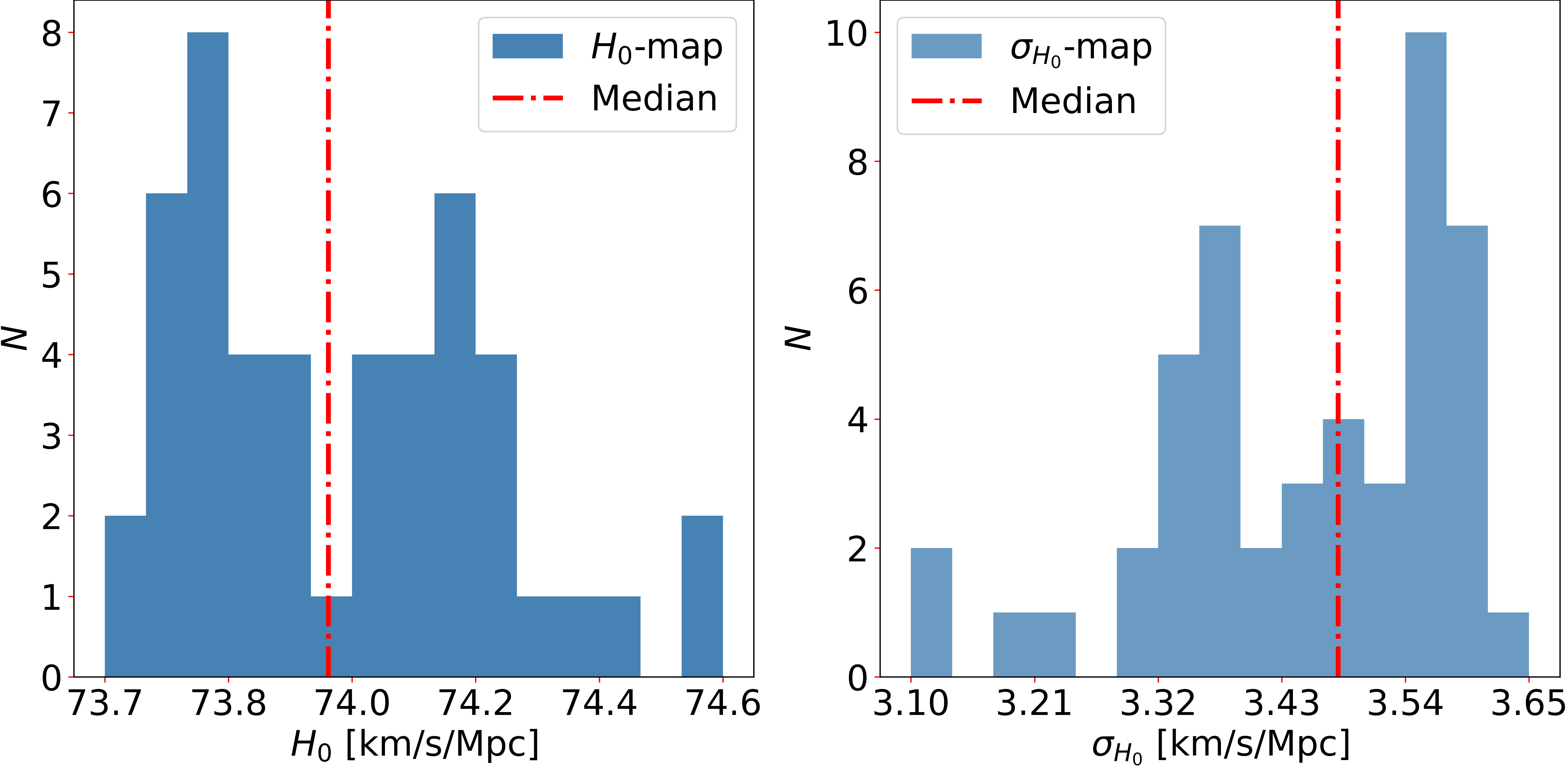} \\
\includegraphics[width=\columnwidth]{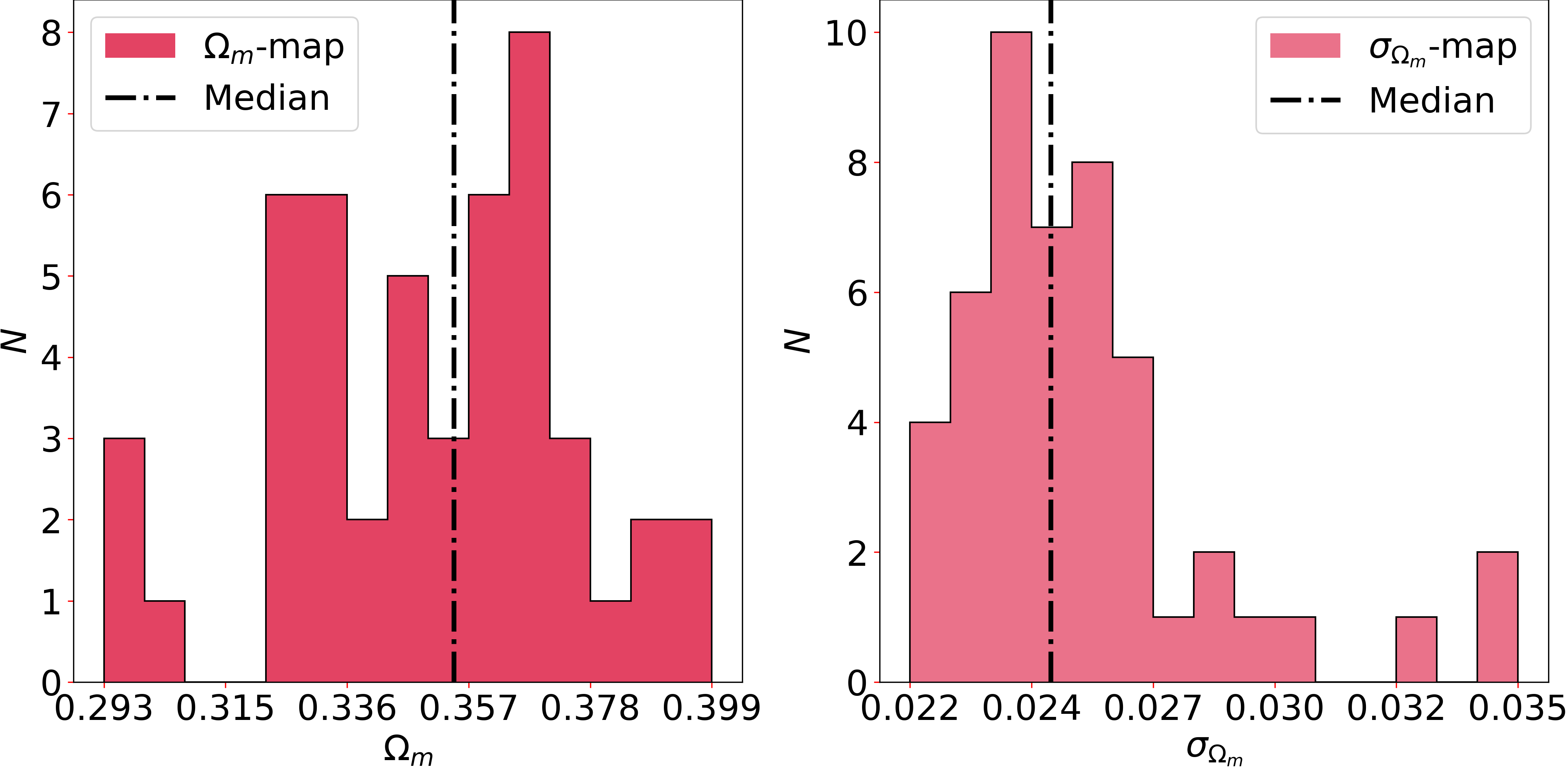} \\
\includegraphics[width=\columnwidth]{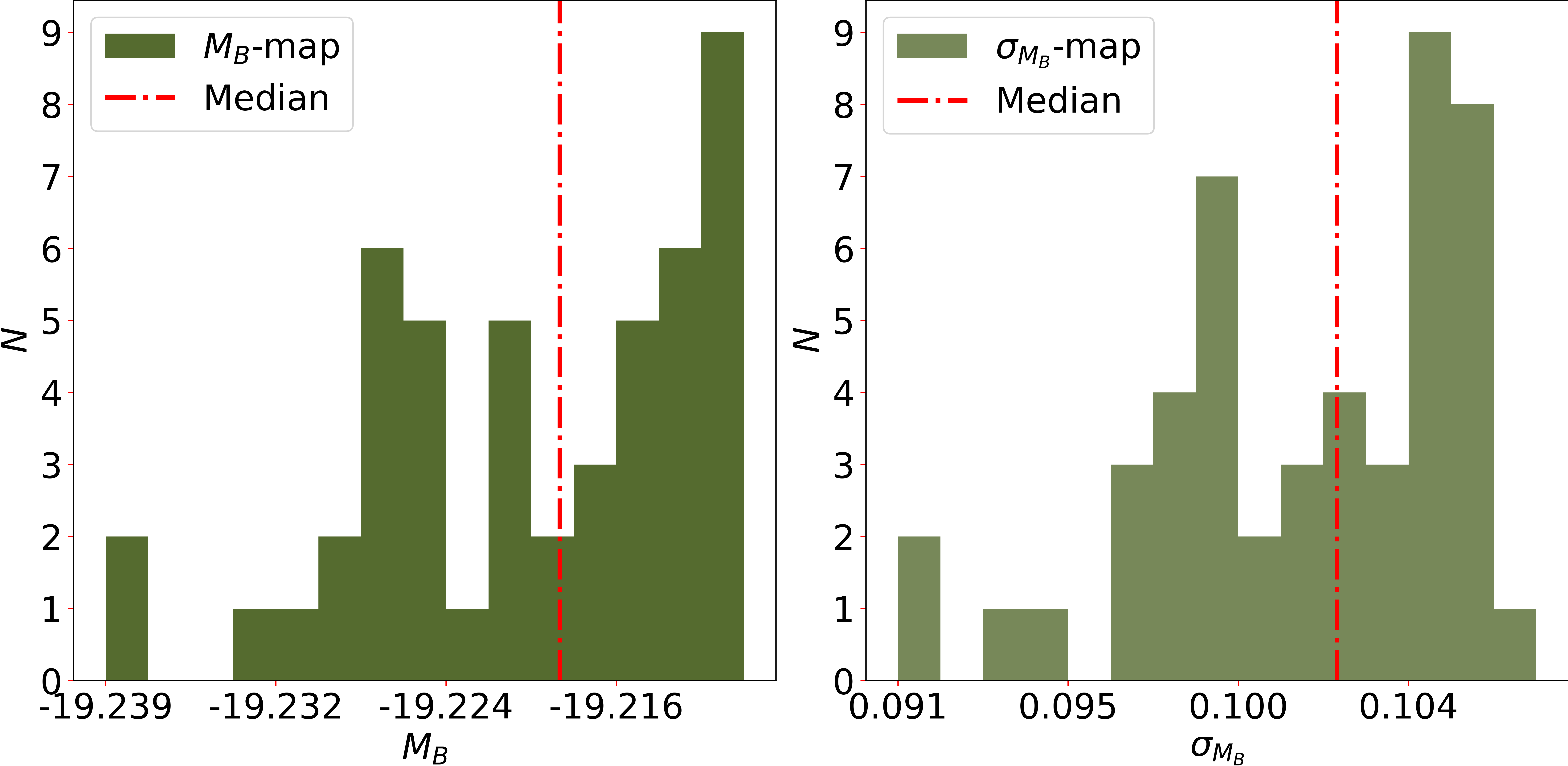}
\caption{
Statistics of the information contained in the maps displayed in Figure~\ref{fig:h0_map_0p015}. 
In the first row, we display the distributions of the pixel values for the $H_0$-map and 
$\sigma_{H_0}$-map, obtaining the medians $73.99$ and $3.48$, and standard deviations $0.24$ and $0.13$ for the $H_0$- and $\sigma_{H_0}$-maps, respectively. 
In the second row, we show the distributions of the pixel values for the $\Omega_m$-map and $\sigma_{\Omega_m}$-map, obtaining the medians $0.354$ and $0.025$, and standard 
deviations $0.025$ and $0.003$ for the $\Omega_m$- and $\sigma_{\Omega_m}$-maps, respectively. 
Instead, in the third row we present the statistics of the 
$M_B$-map and $\sigma_{M_B}$-map, with median values $M_B=-19.22$ with standard deviation $0.007$ and $\sigma_{M_B} = 0.10$ and standard deviation $0.004$; 
one notices that the dispersion of values of this parameter is, indeed, very small.}
\label{fig:gaussian-results}
\end{figure}

The results of the directional analysis of $H_0$ and  $\Omega_m$ across the sky, i.e., the $H_0$-map and the $\Omega_m$-map, are presented in  Figure~\ref{fig:h0_map_0p015}
(and their corresponding uncertainties 
in Figure~\ref{fig:sigmas_map_48}) 
and Tables~\ref{tab:h0_median} and ~\ref{tab:Omega_median} in Galactic coordinates. 
These maps, centered on the Milky Way, illustrate the angular variations of these parameters and were constructed using the HEALPix framework~\citep{Gorski:2004by} for efficient spherical projection.
One observes the existence of a net dipolar pattern in both maps, where the variation between the maximum and minimum values 
are $\sim 1.2 \%$ in the $H_0$-map and  $\sim 26.6 \%$ in the $\Omega_m$-map. This can be seen in Figure~\ref{fig:gaussian-results}, where histograms for the best fits of all free parameters and their corresponding 1$\sigma$ errors in all 48 hemispheres are included. 
In all cases, the dashed lines indicate the median. 
The values of the medians for $H_0$ and $\Omega_m$ are summarized in Tables~\ref{tab:h0_median} and ~\ref{tab:Omega_median}. 
These results can be compared to an MCMC analysis considering the full-sky data of the complete Pantheon+SH0ES catalog 
where one obtains: 
$H_0=73.40 \pm 1.02$, $\Omega_m=0.33 \pm 0.02$, and $M_B=-19.25 \pm 0.03$. 
On the other hand, an analysis of the 
full-sky analysis but considering the redshift interval with $z_{min}=0.015$ gives: $H_0=74.02 \pm 3.24$, $\Omega_m=0.33 \pm 0.02$, and $M_B=-19.23 \pm 0.09$. 
We observe that the full-sky results agree well, at $1\,\sigma$ confidence level (CL), with the directional analysis outcomes shown in Tables~\ref{tab:h0_median} and~\ref{tab:Omega_median}, 
but the full-sky average has a slightly larger impact on the 
matter density value. 


In the full-sky analysis above, for $z_{\min}=0.015$, we note that the uncertainties in $M_B$ increase compared to complete redshift sample. 
This happens because the very low-redshift (and precise) Cepheid calibrators are removed by this redshift cut. 
However, note that in our directional analyses the calibrated SNe are always included whenever present in a given hemisphere, so the calibration is consistently propagated; the broader error bars simply reflect the reduced statistical weight of the calibrators after the cut.

It is also worth noting that part of the apparent $H_0$ variations reported 
in the literature may in fact reflect differences in $M_B$ driven by the 
Cepheid-calibrated SNe at very low redshift, where they are more susceptible 
to peculiar velocity effects. 
Applying a cut at $z_{\min}=0.015$ reduces their impact while still allowing for a robust test of preferred directions of $H_0$ at larger scales.


The complete evaluation of the statistical significance of the $H_0$- and $\Omega_m$-maps angular features is done by analyzing their angular power spectra and performing a comparison with the spectra obtained from the $\{ H_0^{\text{ISO}} \}$-maps and the 
$\{ \Omega_m^{\text{ISO}} \}$-maps, respectively.

\begin{table}[htbp]
\begin{center}
\begin{tabular}{l c c c c} 
\hline
$z_{\text{min}}$ & $H_0^{\text{monopole}}$ & $[H_0^{\text{dip-min}},
H_0^{\text{dip-max}}]$ &\textbf{$l$ ($^{\circ}$)} & \textbf{$b$ ($^{\circ}$)} \\ 
\hline
$0.010$ & 73.175 & $[-1.147,1.147]$ & $303.08$ & $53.46$ \\
$0.015$ & 74.020 & $[-0.346,0.346]$ & $296.96$ & $26.06$ \\
$0.020$ & 74.107 & $[-0.302,0.302]$ & $299.06$ & $24.46$ \\
\hline
\end{tabular}
\end{center}
\caption{Statistical features of the $H_0$-maps for the three cases of 
$z_{\text{min}}$ in analysis: 
Monopole, and dipole limits plus the dipole directions, in galactic 
coordinates, of the $H_0$-maps and their corresponding dipole maps shown in Figure~\ref{fig:h0_map_dipole} in 
Appendix~\ref{z-min-cases}.
}
\label{tab:h0_median}
\end{table}

\begin{table}[htbp]
\begin{center}
\begin{tabular}{l c c c c} 
\hline
$z_{\text{min}}$ & $\Omega_m^{\text{monopole}}$ & $[\Omega^{\text{dip-min}}_m,\Omega^{\text{dip-max}}_m]$ &\textbf{$l$ ($^{\circ}$)} & \textbf{$b$ ($^{\circ}$)}\\ 
\hline
$0.010$ & 0.3486 & $[-0.0367, 0.0367]$ & $105.60$  & $-31.70$ \\
$0.015$ & 0.3495 & $[-0.0354, 0.0354]$ & $104.08$ & $-31.20$ \\
$0.020$ & 0.3430 & $[-0.0321, 0.0321]$ & $107.26$  & $-29.93$ \\
\hline
\end{tabular}
\end{center}
\caption{Statistical features of the $\Omega_m$-maps for the three cases of $z_{\text{min}}$ in analysis: 
Monopole, and dipole limits plus the dipole directions, in galactic coordinates, of the $\Omega_m$-maps and their corresponding dipole maps shown in 
Figure~\ref{fig:Om_map_dipole} in Appendix~\ref{z-min-cases}.
} 
\label{tab:Omega_median}
\end{table}



\begin{figure}[htbp]
\centering
\includegraphics[width=0.8\columnwidth]{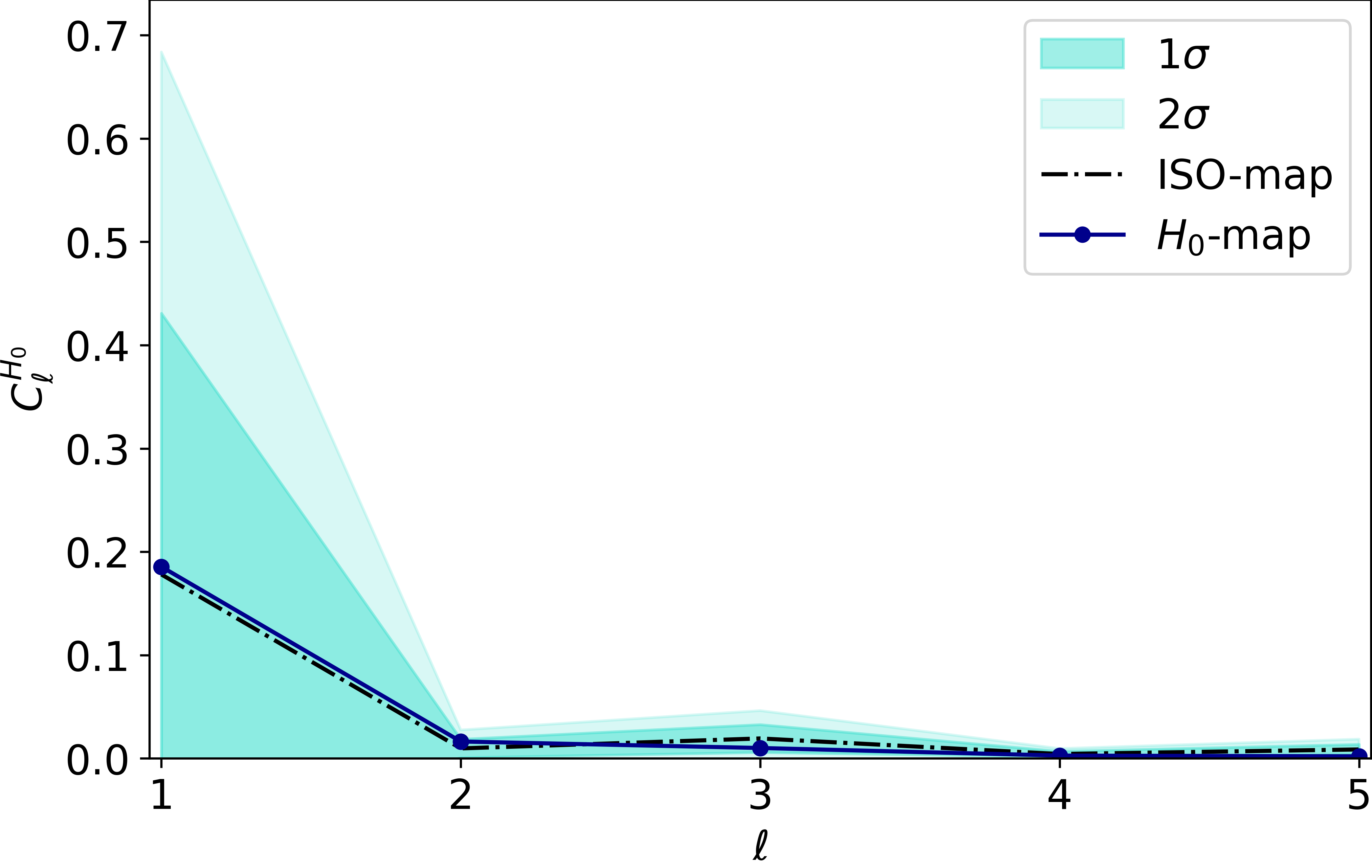}
\includegraphics[width=0.8\columnwidth]{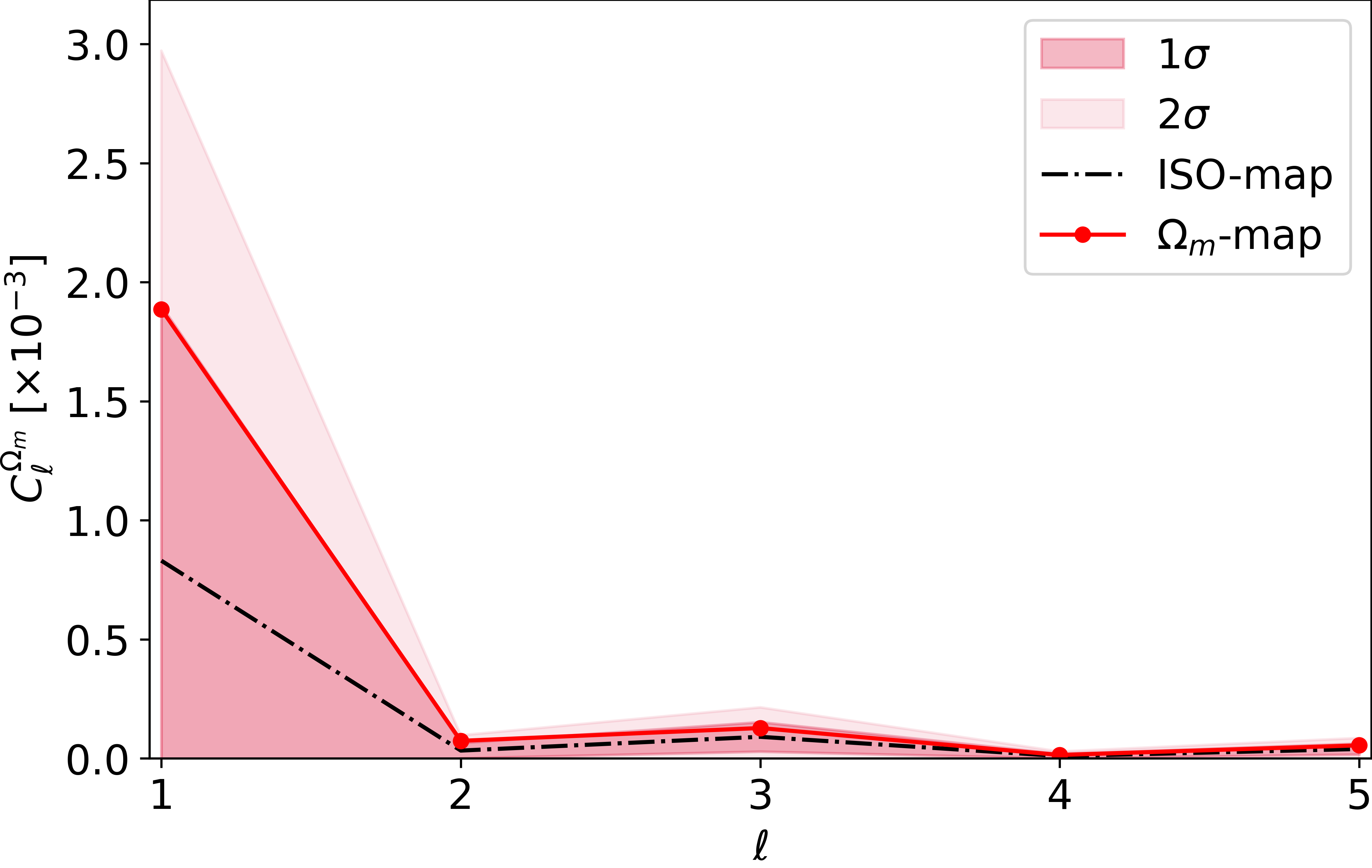}
\caption{The plots illustrate the angular power spectra of the $H_0$ and $\Omega_m$ maps alongside their corresponding ISO-maps. 
The shaded regions represent the 1$\sigma$ and 2$\sigma$ confidence intervals obtained from the ensemble of 1000 ISO-maps. 
The comparison reveals how the observed maps deviate from statistical isotropy, providing insights into possible directional dependencies in $H_0$ and $\Omega_m$. 
These analyses correspond to the SNe sample with $z_{min} \gtrsim 0.015$.} 
\label{fig:iso_test2}
\end{figure}

In fact, in Figure~\ref{fig:iso_test2} we present two plots with the angular power spectra  of the data maps (solid lines), already shown in Figure~\ref{fig:h0_map_0p015}, along with the median power spectrum (dashed lines) and the 1$\sigma$ and 2$\sigma$ regions (colored regions) 
for $1000$ $\{ H_0^{\text{ISO}} \}$-maps and 
$1000$ $\{ \Omega_m^{\text{ISO}} \}$-maps, 
respectively. 
From this analysis, one concludes that the dipolar behavior in the $H_0$- and $\Omega_m$-maps exhibits no significant anisotropy at large scales. Actually, one observes that the power spectrum of the data maps lies within the $1\,\sigma$ region in both cases, namely the $H_0$-map and for the 
$\Omega_m$-map, reinforcing the expectation of an isotropic expansion and matter distribution in the universe. 
In this sense, the absence of a significant directional preference in both the $H_0$- and $\Omega_m$-maps suggests that our results are consistent  with the flat-$\Lambda$CDM concordance model in describing the angular distribution of matter density across the universe. 
These results agree with previous studies that investigated possible preferred directions of cosmological parameters using SNeIa data and other cosmic probes (see, e.g.,~\cite{Hu20,conville2023,tang2023,Wang23,Wu25}).

To put our findings into context, it is important to compare the measured dipolar preferred directions with previous determinations of large-scale anisotropic signals. The CMB dipole was determined by Planck  \citep{Planck2018:Aghanim} as  $(l,b) = (264.021^\circ \pm 0.011^\circ
, 48.253^\circ \pm 0.005^\circ
)$. Local peculiar velocity studies, such as  \cite{Qin2018}, report a bulk flow on $< 50 h^{-1}$ Mpc scales toward $(l,b) \sim (300^\circ,23^\circ)$. On larger scales, \cite{Watkins2023} find a bulk flow extending to $\sim 200 h^{-1}$ Mpc toward $(l,b) \sim (298^\circ,-8^\circ)$. On  much larger scales, it has been suggested a coherent motion often referred to as {\em dark flow}. 
For instance, \cite{Kashlinsky2010} used the kinematic Sunyaev–Zel’dovich effect to measure a flow in the direction  $\sim (283^\circ,20^\circ)$. Moreover, \cite{Mariano2012}, analyzing the Keck+VLT quasar absorber and  Union2 Sne Ia samples, reported a dipole direction at $(l,b)=(309^\circ,-15^\circ)$, while  \cite{Migkas2021}, studying the X-ray luminosity–temperature relation of galaxy clusters, found a significant anisotropy with a preferred dipole direction at $(l,b)=(280^\circ,-15^\circ)$. 
Our results, $(l,b)=(303.1^\circ,53.5^\circ)$ at $z \sim 0.01$, lie in the same broad Galactic longitude range as most previous determinations but at a higher Galactic latitude, indicating longitudinal consistency but latitude-dependent variations. 
However, these comparisons must be treated with caution, as preferred directions depend strongly on redshift and sky coverage analyzed. 
For example, at $z=0.01$ which corresponds to $\sim \!40,h^{-1}$ Mpc, i.e. the very local universe, where cosmic variance and peculiar velocities dominate the analyses~\citep{Tully_2023}.

\begin{figure}[htbp]
\centering
\includegraphics[width=0.8\columnwidth]{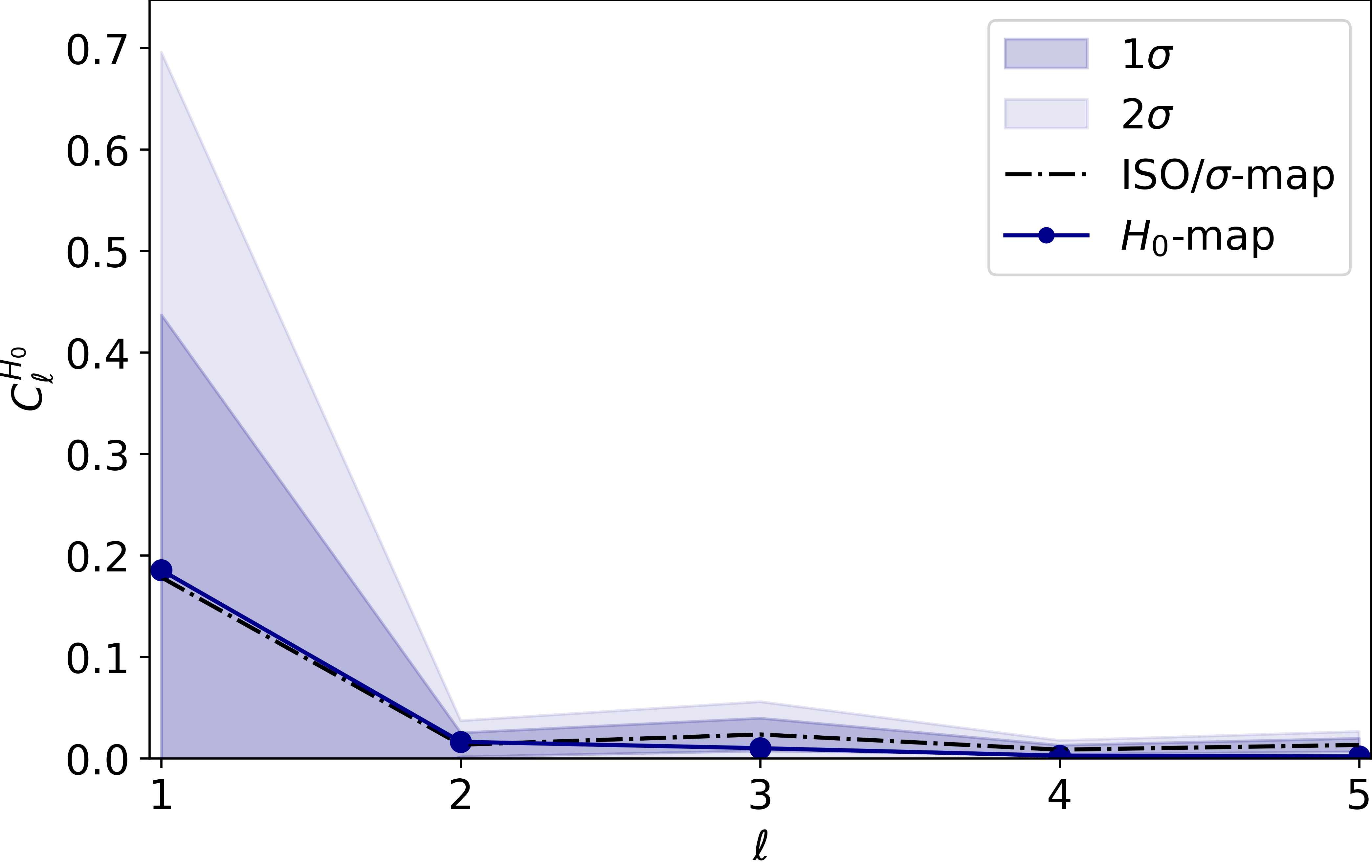}
\includegraphics[width=0.8\columnwidth]{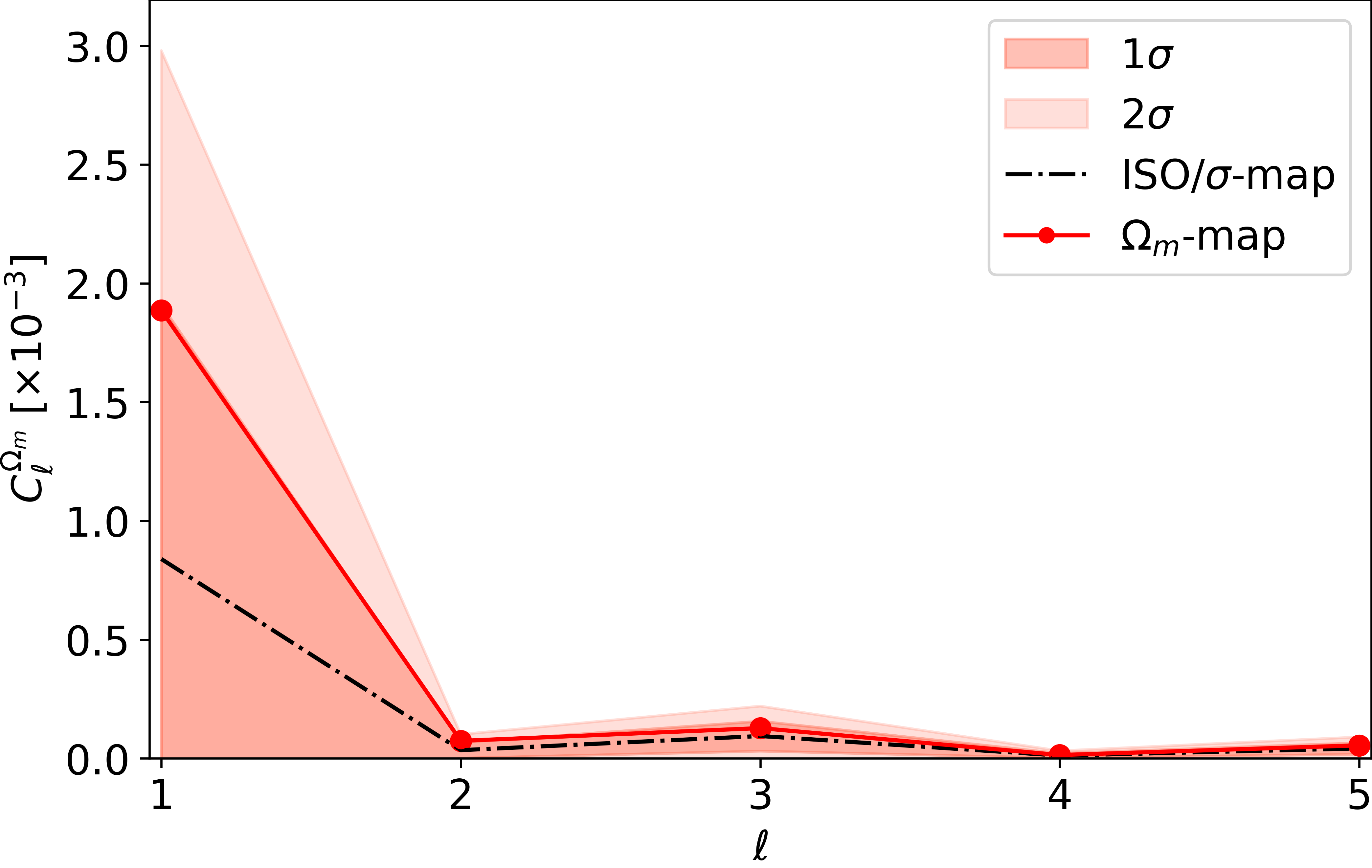}

\caption{
The angular power spectra of the 
$H_0$ and $\Omega_m$ maps compared with 
the power spectra of $1000$ ISO/$\sigma$-maps, that is 
to the original set of ISO-maps we have randomly added 
the observational uncertainties calculated in our best-fit 
procedure and shown in the $\sigma_{H_0}$- and 
$\sigma_{\Omega_m}$-maps in Figure~\ref{fig:sigmas_map_48}. 
The shaded regions represent the $1\sigma$ and $2\sigma$ 
confidence intervals.
}
\label{fig:sigma_consistency}
\end{figure}

We also calculate the uncertainties associated with the parameters $H_0$ and $\Omega_m$ obtained from the best-fit values, and 
assemble them as full-sky maps, termed the 
$\sigma_{H_0}$- and $\sigma_{\Omega_m}$-maps, 
displayed in Figure~\ref{fig:sigmas_map_48}. 
We are interested in studying the impact of these uncertainties on the observed dipolar pattern of the $H_0$- and $\Omega_m$-maps. 
For this reason, we perform a consistency test as follows: we added the ISO-maps (see Section \ref{isotropic-maps}) to the maps of parameter uncertainties (see Figure \ref{fig:sigmas_map_48}), the $\sigma_{H_{0}/\Omega_{m}}$-maps, in a shuffled manner, i.e.,  ISO/$\sigma$-map$^i$=ISO-map$^{i} + \sigma$-map$^{i}_{\text{shuffle}}$, for $i = 1, \dots, 1000$. This approach  preserves the statistical distribution of the uncertainties while removing possible directional correlations. 
From the set of 1000  simulated  ISO/$\sigma$-maps, we calculated the angular power spectra for each realization, along with their median and the 1$\sigma$ and  2$\sigma$ regions, and compared them with the observed data,  in  Figure \ref{fig:sigma_consistency}. We conclude that the uncertainties in the parameters $H_0$ and $\Omega_m$ do not introduce significant changes in the angular power spectrum or its statistical significance, reinforcing the robustness of our results.

%

\begin{figure}[htbp]
\centering
\includegraphics[width=0.78\columnwidth]{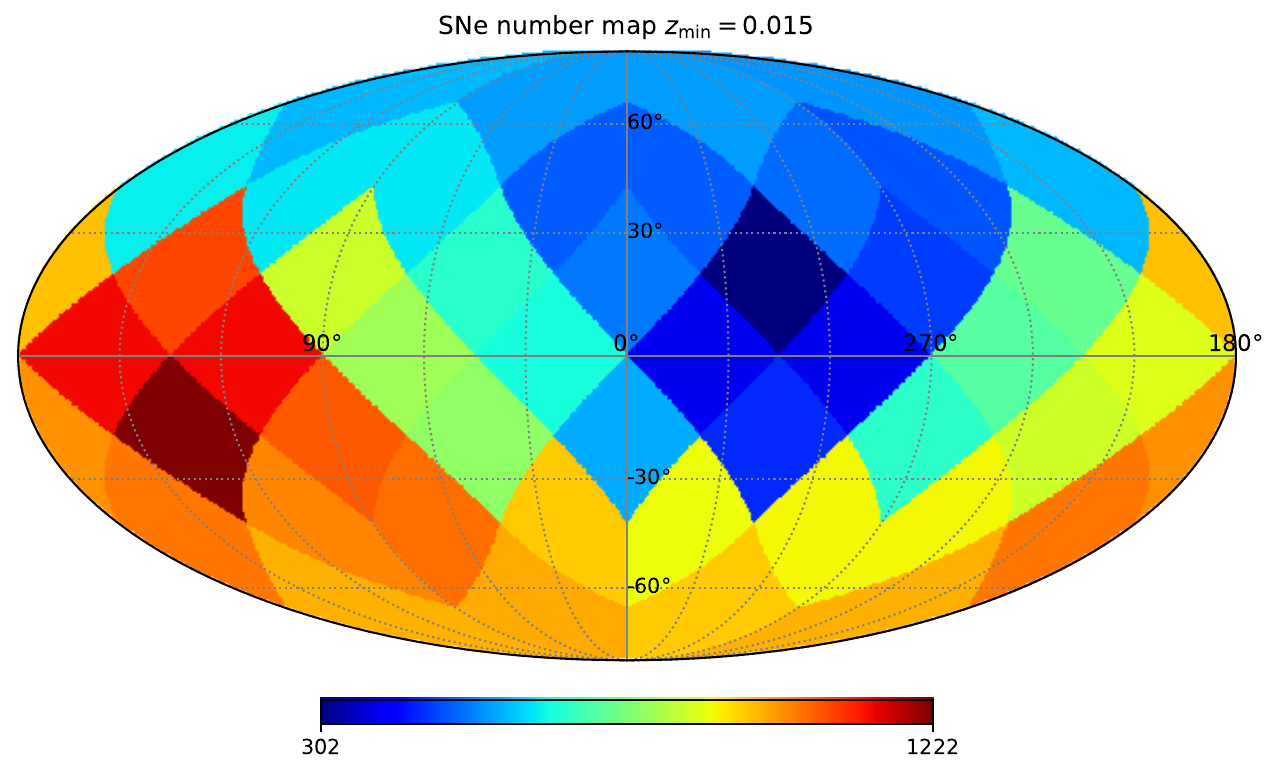}
\caption{Number of SNe-map, or $N$-map, considering 48 hemispheres, for the case $z_{\text{min}}= 0.015$, in Galactic coordinates. It provides the number of SNe analyzed in each of the 48 hemispheres defined in our directional analyses of the $H_0$ and $\Omega_m$ parameters.
}
\label{fig:num_sn_0015}
\end{figure}

Complementing this analysis, we also investigate the possibility that the dipolar pattern could be an effect related to the number of SNe in the hemispheres, $\{ n_J \}$, that is, the Number of SNe-map (N-map), shown in Figure~\ref{fig:num_sn_0015}. We consider this possibility due to the correlation between the $H_0$- and $\Omega_m$-maps and the $N$-map, as revealed by the Pearson correlation coefficient. 
The Pearson coefficient quantifies the strength of linear correlations between two 
variables, with values in the range $[-1,1]$. 
For clarity, we classify correlations according to the absolute value of this coefficient: 
 
\begin{itemize}
\item 
\textbf{$0.0 \le |C| < 0.2$: very weak or no correlation;} 
\item 
\textbf{$0.2 \le |C| < 0.4$: weak correlation;} 
\item 
\textbf{$0.4 \le |C| < 0.6$: moderate correlation;} 
\item 
\textbf{$0.6 \le |C| < 0.8$: strong correlation;} 
\item 
\textbf{$0.8 \le |C| \le 1.0$: very strong correlation.} 
\end{itemize}

In our case, Corr($N$-map, $H_0$-map) =  $-0.799$ ($|C| \sim 0.8$) and Corr($N$-map, $\Omega_m$-map) = $0.703$ ($|C| \sim 0.7$), which correspond to strong anti-correlation 
and strong correlation, respectively, between the analyzed maps. 
For this reason we investigate in detail, 
in ~\ref{supernova-number-analysis}, 
the impact of the number of SNe on our directional analysis of $H_0$.

Lastly, and equally important, the correlation between 
the $H_0$-map and the $\Omega_m$-map is 
Corr($H_0$-map,$\Omega_m$-map) = -0.914, 
which is expected because these parameters are 
inversely proportional, as observed in equation~(\ref{H(z)}) 
in the flat-$\Lambda$CDM model. 


Recently,~\cite{perivolaropoulos2023} uses hemispherical analyses to study deviations of isotropy for the absolute magnitude in the Pantheon+SH0ES sample, finding consistency with simulated Monte Carlo catalogs for different redshift bins. However, for redshift bins of data with distances below 40 Mpc, a sharp change in anisotropy is detected. 
As we observe in the third group histogram of Figure~\ref{fig:gaussian-results}, the results of our directional analysis of $M_B$ for $z_{min}= 0.015 \,\,  (\sim 60 ~\text{Mpc})$ show, indeed, a tiny dispersion of values across the sky, 
of $(0.007/19.22)\times 100 \simeq 0.04\%$ around the median value, compatible with the 
result obtained by~\cite{perivolaropoulos2023}.

While both $H_0$ and $\Omega_m$ can show apparent anisotropies in supernovae samples, the underlying systematics differ. 
For $H_0$, peculiar velocities remain the dominant concern at low redshift, whereas $\Omega_m$ is more vulnerable to residual differences among the small compilations --from diverse surveys-- that compose the data organized in the Pantheon+SHOES sample. 
As highlighted in earlier works 
(see, e.g.,~\citep{Scolnic_2018,brout2022, Malekjani2024}), 
certain systematics can cause redshift-dependent trends, which, translated into variations in cosmological parameters, could eventually appear as directional variations in the sky. 
This caution is important since $\Omega_m$ is a parameter dependent on the 
cosmological model.

Our results, summarized in Table~\ref{tab:Omega_median}, 
indicate that the monopole values, as well as the dipole range values 
and the directions obtained are very stable in the three cases analyzed, i.e., $z_{min} = 0.01, 0.015, 0.02$. 
In this sense, we interpret the directional variations of $\Omega_m$ 
as generated by the distribution of matter, projected onto the sky, 
from the three-dimensional volumes for each redshift bin in analysis.


In addition to the directional analysis discussed in this section, in Appendices~\ref{app:high_reso} and ~\ref{z-min-cases}, we present the results of various robustness and consistency tests, investigating higher angular resolution maps (with 192 hemispheres) and samples of SNe with  $z_{min}= 0.01$ and  $0.02$. 
In Figures~\ref{fig:iso_z001} and \ref{fig:iso_z002} (see Appendix~\ref{z-min-cases}) we present the results corresponding to redshifts cases $z_{min} = 0.01$ and $z_{min} = 0.02$, respectively. 
In the former case, we observe a large dipole component for the 
$H_0$-map, at more than $2\,\sigma$ CL, suggesting a violation of isotropy in the angular distribution of the $H_0$ values. 
However, this high dipole component diminishes to value within 
$1\,\sigma$ CL in the analyses of the samples with 
$z_{min} = 0.015$ and $z_{min} = 0.02$, as observed in the Figures~\ref{fig:iso_test2} and~\ref{fig:iso_z002}, respectively. 
This suggests that, for $z_{min} = 0.01$, the observed anisotropy can be attributed primarily to local effects, which progressively diminish as $z_{min}$ increases and isotropy is recovered for $z_{min} \gtrsim 0.015$. 
Although this behavior manifests in both cosmological parameters, $H_0$ and $\Omega_m$, the influence of peculiar velocities directly impacts the estimation of~$H_0$. 
%

\section{Conclusions and Final Remarks}\label{Summary and conclusions}

In this study, we investigated the angular distribution of the Hubble constant ($H_0$) and the matter density 
($\Omega_{m}$) across the sky using the Pantheon+SH0ES Type Ia supernovae catalog. 
Through a hemispherical analyses described in section~\ref{directional}, we constructed maps with directional information of these cosmological parameters and analyzed their statistical significance within the framework of the flat-$\Lambda$CDM model and in the CMB frame. 
We analyzed the sample of SNe with redshift $z \in [0.015, 2.261]$, 
i.e., $z_{\text{min}} = 0.015$. 
Our results show the existence of dominant dipoles in the $H_0$-map and $\Omega_m$-map, although both consistent with statistical isotropy within $1\sigma$ CL 
for $z_{\text{min}} \gtrsim 0.015$ (i.e., $\sim \!60$ Mpc). 
However, for nearby SNe --at distances 
$\lesssim 60$ Mpc-- our consistency analyses in Appendix~\ref{z-min-cases} show that peculiar velocities introduce a highly significant dipole in the angular distribution of $H_0$. 

In fact, in Appendix~\ref{z-min-cases}, we study the $H_0$- and 
the $\Omega_m$-maps for SNe samples with other $z_{\text{min}}$ values, obtaining the corresponding dipole directions (shown in the maps displayed in Figures~\ref{fig:h0_map_dipole} and~\ref{fig:Om_map_dipole}), 
their statistical significance analyses (displayed in Figures~\ref{fig:iso_z001} 
and \ref{fig:iso_z002}), and complementary information given in 
tables~\ref{tab:h0_median} and~\ref{tab:Omega_median}. 
This directional analysis of the $H_0$- and the 
$\Omega_m$-maps for the different cases of $z_{\text{min}}$ illustrates the impact of  low-$z$ data on the statistical significance of the dipolar pattern of the parameters maps, an effect likely caused by large peculiar velocities in the Local Universe~\citep{Avila2023,Lopes2024,Sorrenti24,Sorrenti2024c,Swati2025,Courtois2023,Courtois2025,Marinoni2023}.

For the $\Omega_m$-map analysis, one observes that 
the lack of significant anisotropy observed in Figure~\ref{fig:iso_test2}, contrasts with some studies, such as \cite{Javanmardi2015}, who reported isotropy violation in matter density parameter using alternative datasets (although, caution is needed for comparisons involving different data sets). Our results, instead, confirm the isotropic distribution of matter at cosmological scales, adding evidence to support flat-$\Lambda$CDM as the concordance model of cosmology in reproducing features of the observed universe 
(see, e.g.,~\cite{Appleby2014,Marques2018,Avila2022,Lopes2025}). 
While small fluctuations in the distribution of matter are observed, they are consistent with statistical isotropy and do not indicate any significant departure from the predictions of the standard $\Lambda$CDM model. 
Finally, for the intrinsic magnitude of the SNe, $M_B$, we do not find any significant anisotropies for distances larger than $\sim \!60$ Mpc, consistent with~\cite{perivolaropoulos2023}.

It is worth noting that part of the apparent $H_0$ variations reported 
in the literature may in fact reflect differences in $M_B$ driven by the 
Cepheid-calibrated SNe at very low redshift, where they are more susceptible 
to peculiar velocity effects. 
Applying a cut at $z_{min} = 0.01, \,0.015, \,0.02$, reduces their impact while still allowing for a robust test of preferred directions of $H_0$ at larger scales.


Additionally, we also studied the impact on our results coming from uncertainties in $H_0$ and $\Omega_m$ (shown in figure~\ref{fig:sigma_consistency}), the number of SNIa in each region of the sky (in Appendix~\ref{supernova-number-analysis}), and different samples of SNe (in Appendix~\ref{z-min-cases}). 
Our results, in all cases, are quite robust.

In conclusion, our findings are broadly consistent with the $\Lambda$CDM framework, with the observed $H_0$ dipole pattern likely originating from local effects that diminish at higher redshifts. The robustness tests, including isotropic realizations and statistical analyses across hemispheres, further validate our results. Future studies with higher-resolution datasets and alternative cosmological tracers will be essential to disentangle local contributions from genuine cosmological anisotropies and refine our understanding of the universe's large-scale structure. 
Ultimately, we confirm that, based on our analyses and at the current precision of the Pantheon+SH0ES dataset, 
the cosmological principle is valid.


\section*{Acknowledgements}
The authors acknowledge the use of  data from Pantheon+SH$0$ES. 
We also acknowledge the use of the CHE cluster, managed and funded by the COSMO/CBPF/MCTI, with financial support from FINEP and FAPERJ, operating at Javier Magnin Computing Center/CBPF, 
and the CDJPAS high-performance computing cluster at the Observatório Nacional Data Center (CPDON). 
Additionally, we acknowledge the use of the healpy/HEALPix package~\citep{Gorski2005} for processing and analyzing data.
FA thanks to FAPERJ, Processo SEI-260003/001221/2025, for the financial support. 
ML and AB acknowledges to CAPES and CNPq, for their corresponding fellowships. WSHR thanks CNPq and FAPES for their partial financial support.  RM acknowledges the financial support from CNPq under the fellowship Processo 302370/2024-2.


\bibliographystyle{elsarticle-harv} 
\bibliography{bib}

\appendix

\section{Directional analysis with higher angular resolution}\label{app:high_reso}

\begin{figure}[htbp]
\includegraphics[width=\columnwidth]{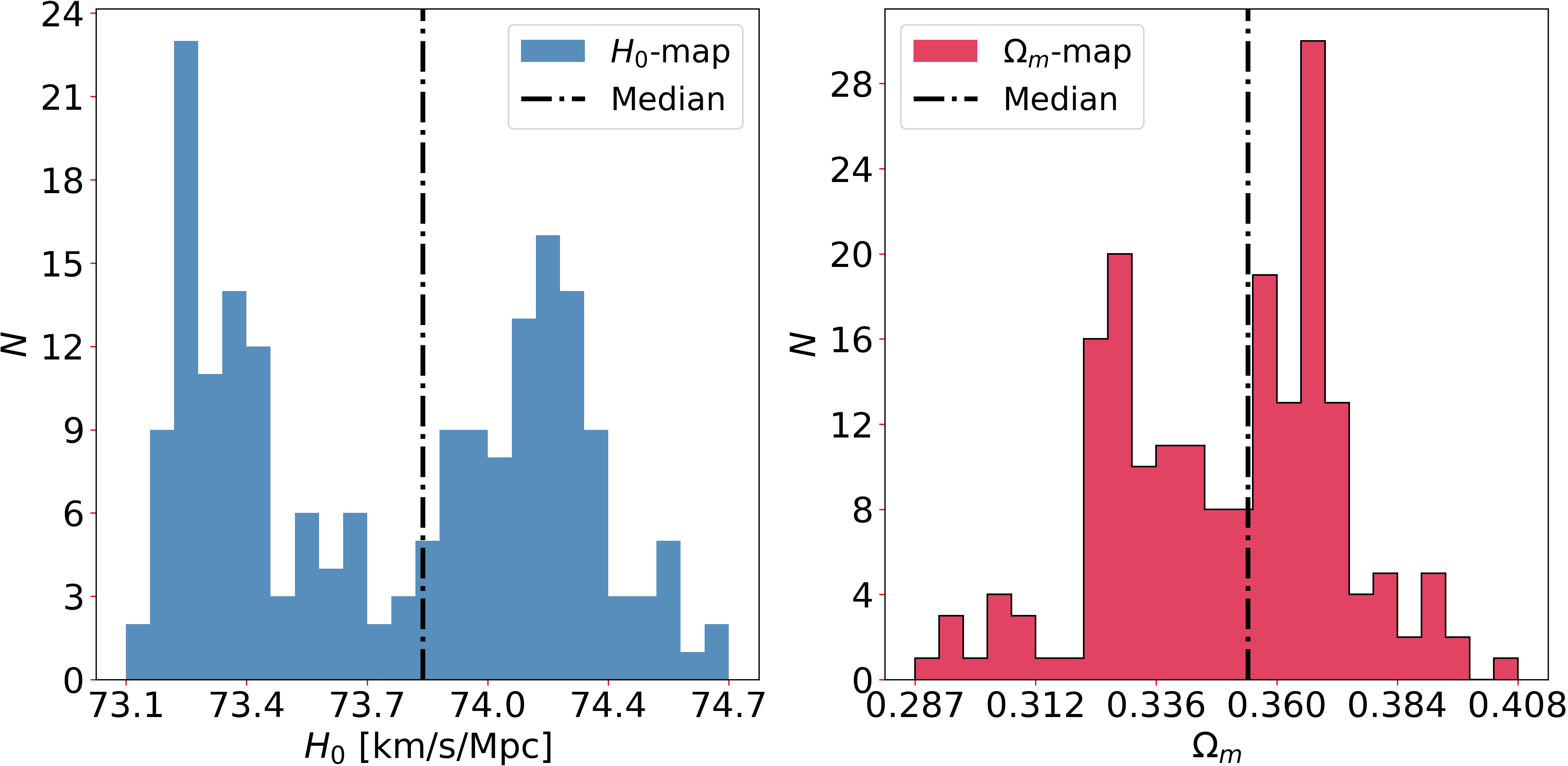}
\caption{
Statistical features of the $H_0$- and $\Omega_m$-maps, shown in 
Figure~\ref{fig:192_maps}, with medians $73.88$ and $0.354$ and 
standard deviations $0.43$ and $0.023$, respectively. 
The analyzed maps contain 192 hemispheres.
}
\label{fig:appA}
\end{figure}

\begin{figure}[htbp]
\centering
\includegraphics[width=0.78\linewidth]{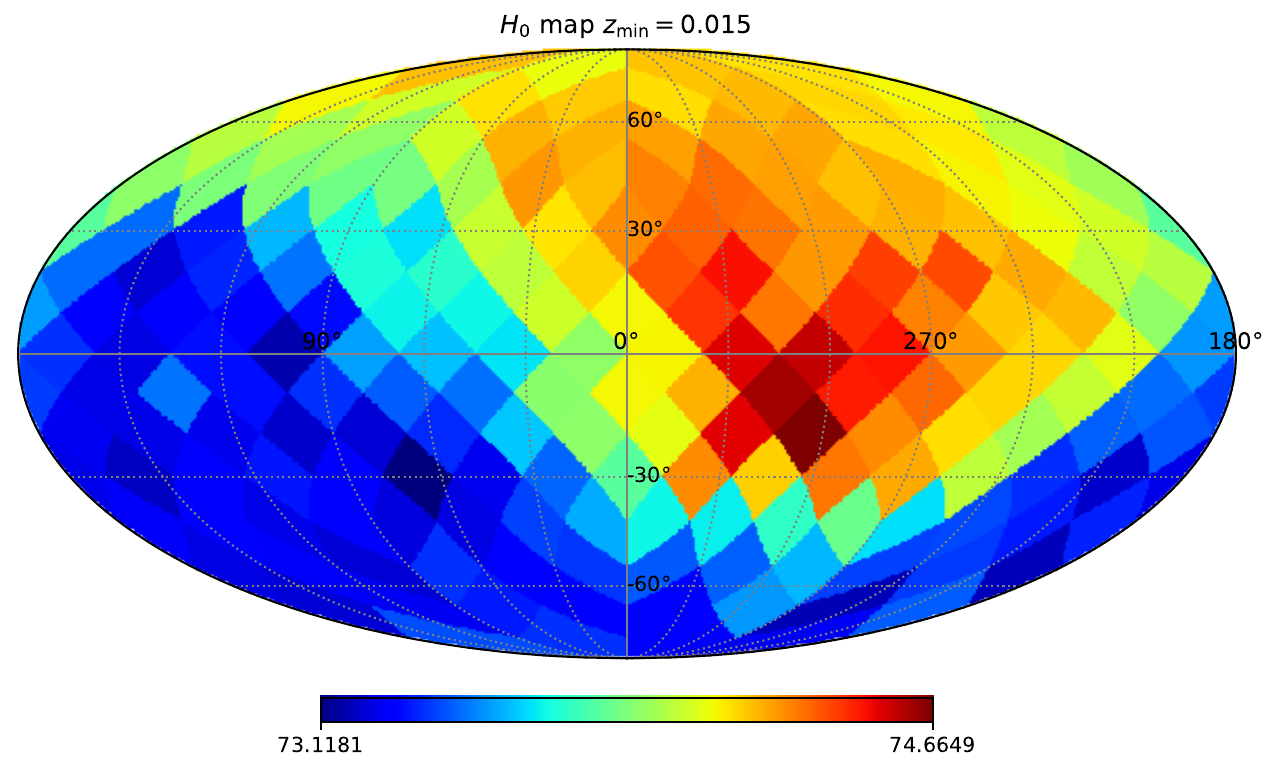}
\includegraphics[width=0.78\linewidth]{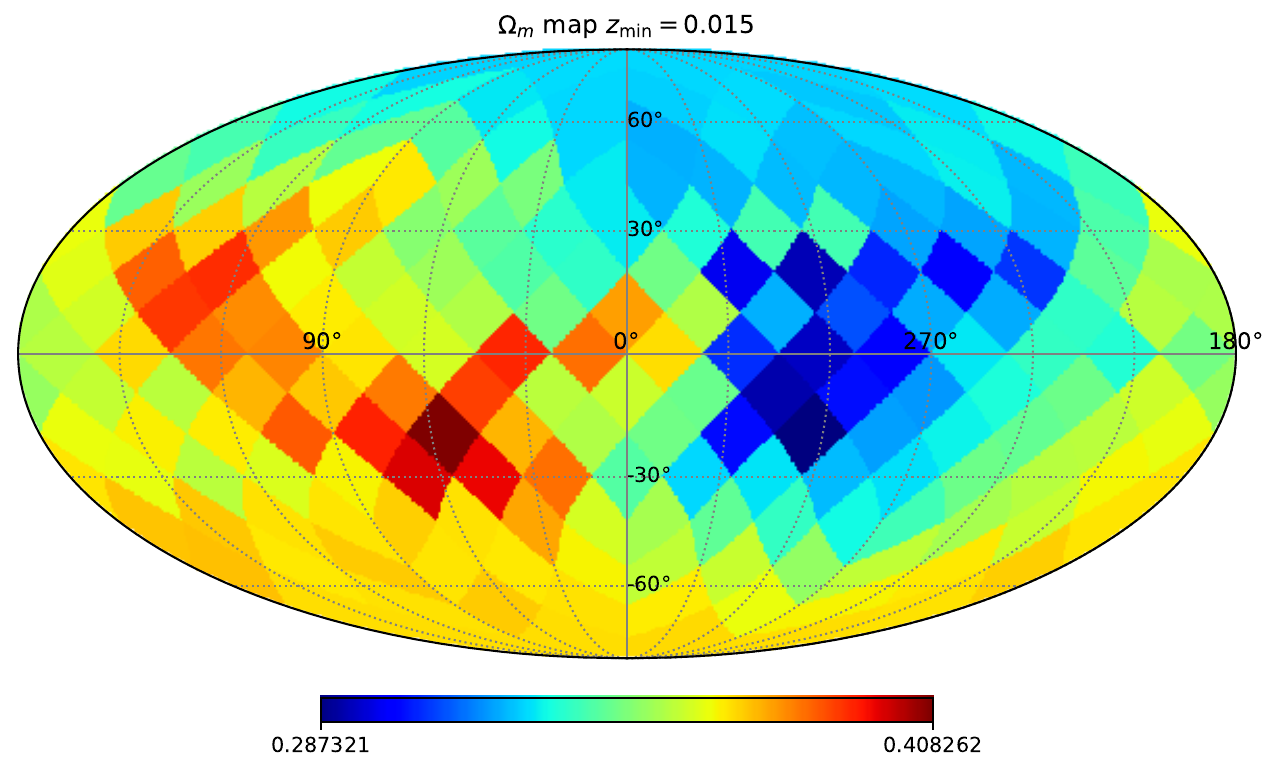}
\caption{
The $H_0$-map (upper map) and $\Omega_m$-map (lower map) produced 
analyzing 192 hemispheres and considering the case 
$z_{\text{min}} = 0.015$. 
}
\label{fig:192_maps}
\end{figure}

In this Appendix, we present consistency results by constructing $H_0$ and 
$\Omega_m$ maps at a higher angular resolution, that is, considering 192 hemispheres, and discussing the case with $z_{\text{min}}=0.015$ (the same studied in 
Section~\ref{results} using 48 hemispheres). The statistical results of the $H_0$- and the $\Omega_m$-maps 
are shown in the histograms displayed in 
Figure~\ref{fig:appA}. 
While in Figure~\ref{fig:192_maps} we show the $H_0$- and the $\Omega_m$-maps 

As one can observe, the results obtained in Section~\ref{results} remain robust.

\section{Impact of the number of supernovae in 
our directional analysis}\label{supernova-number-analysis}

\begin{figure}[htbp]
\centering
\includegraphics[width=0.82\linewidth]{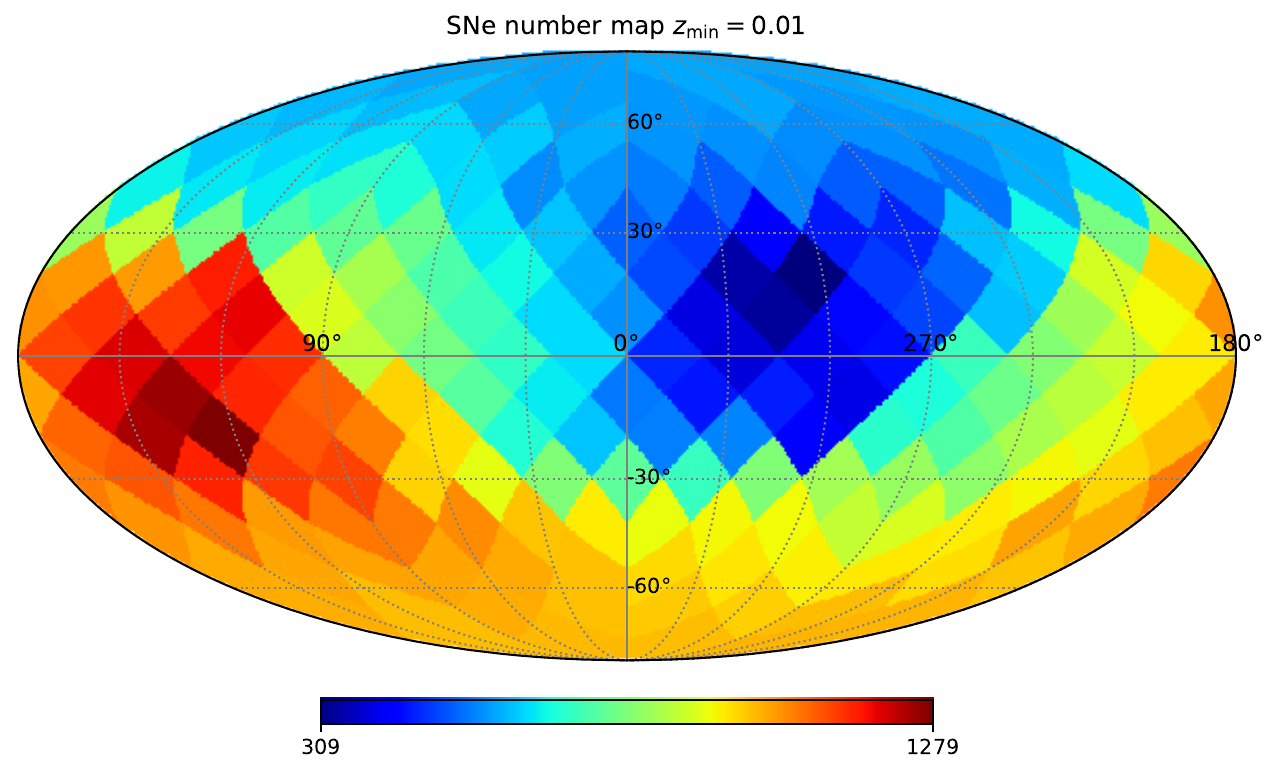}
\includegraphics[width=0.82\linewidth]{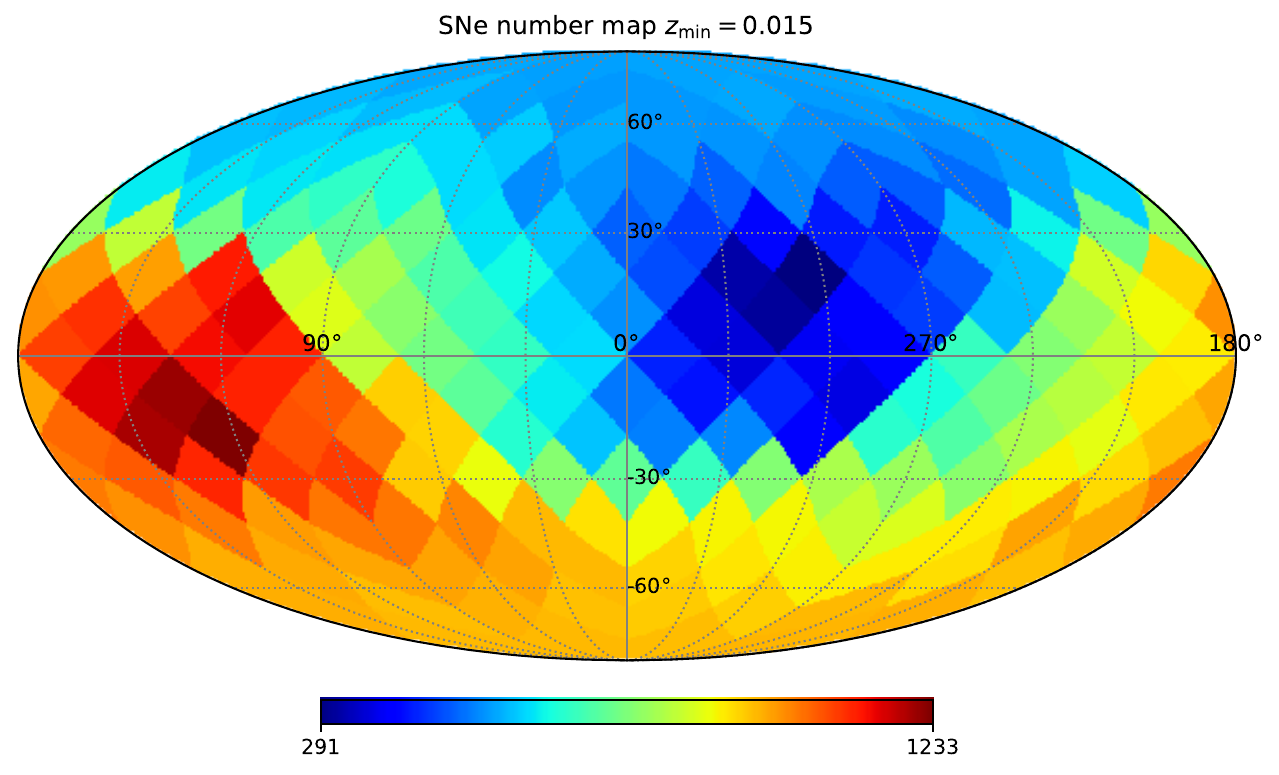}
\includegraphics[width=0.82\linewidth]{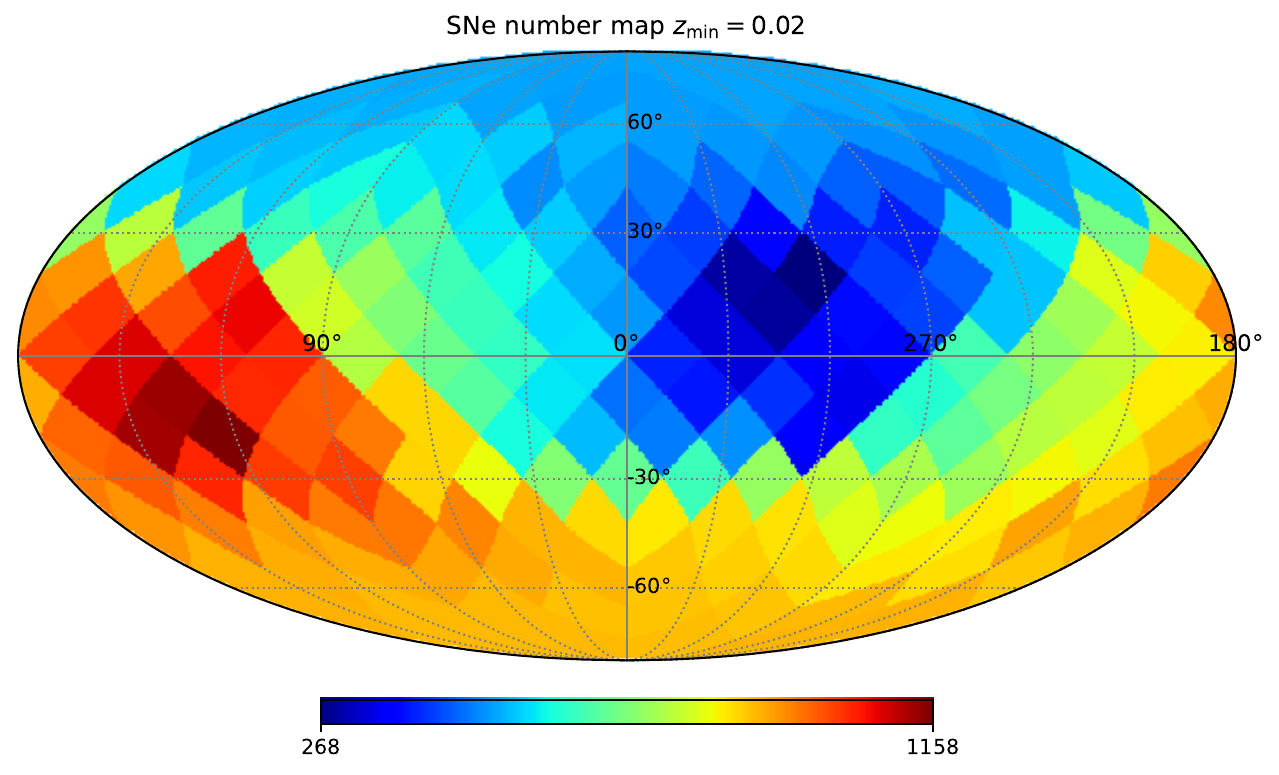}
\caption{From top to bottom: 
Number of SNe-maps, or $N$-maps, considering 192 hemispheres, for the cases $z_{\text{min}}= 0.01$, $z_{\text{min}}= 0.015$, and $z_{\text{min}}= 0.02$.
}
\label{fig:num_sn}
\end{figure}

Observing the Figure~\ref{fig:footprint_SNe}, one clearly notices that the distribution of SNe is not uniform across the sky. 
This leads us to question whether the calculation of the cosmological 
parameters done in our analyses could be biased by the different number 
of SNe in each hemisphere. 
To investigate this, we calculate the Number-of-SNe map, assembled counting 
the number of SNe in each hemisphere, $\{ n_J \}, J=1,2,\cdots, 48$ and 
termed the $N$-map, shown in Figure~\ref{fig:num_sn}, where we display these 
maps for the cases $z_{\text{min}} = 0.01, 0.015, 0.02$. 
The possible negative correlation, or anti-correlation, between this $N$-map and the $H_0$-map is indeed confirmed with the Pearson coefficient, where we obtain the value: $-0.799$. 
This anti-correlation means that in regions containing a smaller  number of SNe, our analyses result in higher values of the parameter $H_0$. 
Our next analyses investigate if this anti-correlation is indeed biasing the dipolar structure of the $H_0$-map. 
These analyses were done for the $z_{\text{min}} = 0.01$ case. 

\begin{figure}[htbp]
\centering
\includegraphics[width=\columnwidth]{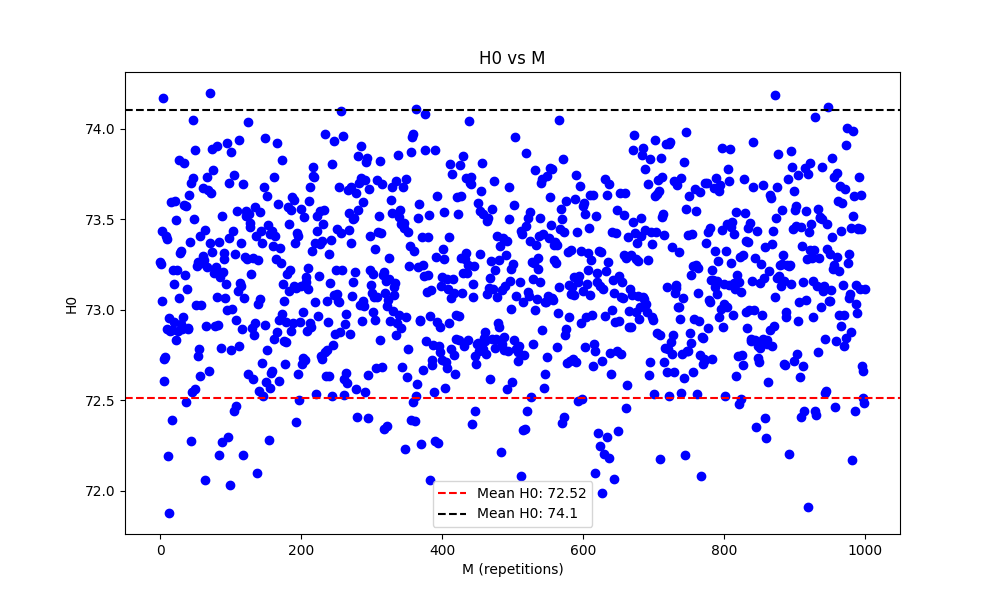}
\includegraphics[width=\columnwidth]{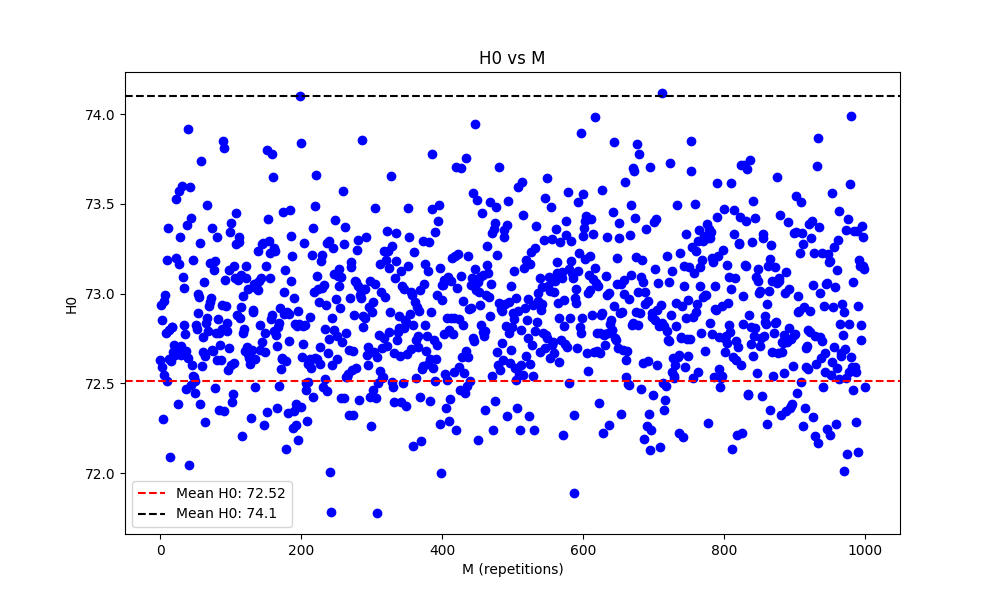}
\includegraphics[width=\columnwidth]{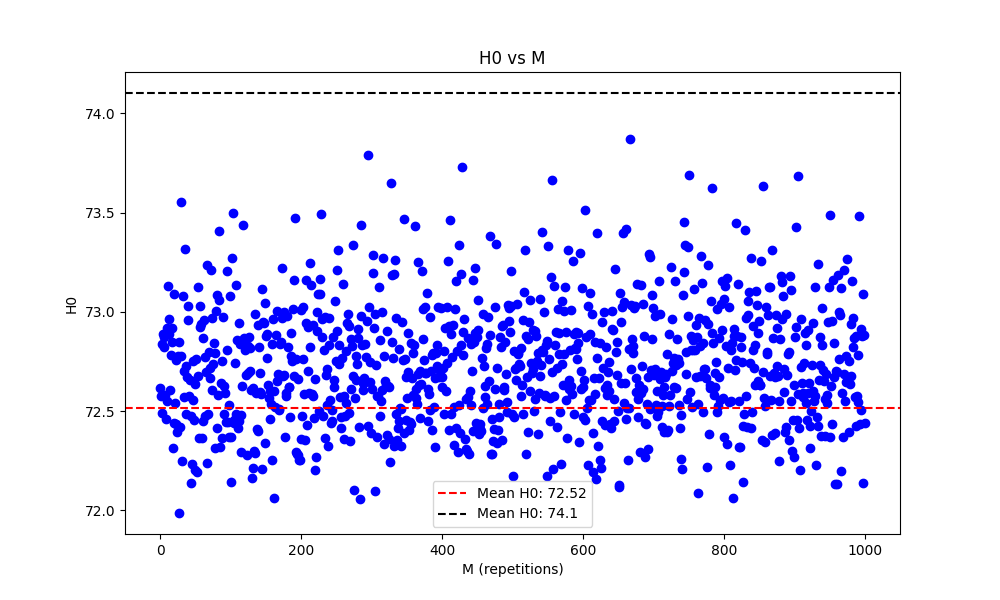}
\caption{
Monte Carlo analyses that calculates $H_0$ in three samples with different number of SNe randomly selected, namely: $309,\,512$, and $700$ SNe 
(displayed in this order from top to bottom). 
See Appendix~\ref{supernova-number-analysis} for details. 
The two dashed horizontal lines indicate the minimum, 
$72.52$ km/s/Mpc, and the maximum $74.1$ km/s/Mpc, values of $H_0$ obtained in the directional analysis of the case $z_{\text{min}}=0.01$ with 192 hemispheres.
}
\label{fig:robustness_test}
\end{figure}
This investigation consists on various robustness tests, based on Monte Carlo analysis, to discover a possible bias in the dipolar direction of the $H_0$-map due to the number of SNe in that direction. 
Specifically, we examine the hemisphere containing the highest number of supernovae, i.e., $1279$ SNe, by randomly selecting  three samples from it, containing $309,\,512$, and $700$ SNe (note that $309$ is the lowest number of SNe obtained in the hemisphere distribution for the case under study, i.e., $z_{\text{min}} = 0.01$, and 192 hemispheres; see the map at the top in Figure~\ref{fig:num_sn}). 
Then we perform a series of Monte Carlo analyses for each 
sample, that is, we repeat the above choice of SNe samples a number 
$\text{M}$ of times, considering 
$\text{M} = 10, 20,\cdots, 1000$,  
and calculate $H_0$ in each case. 

For each set of $\text{M}$ repetitions, we calculate the 
$\bar{H}_0$ median, and then plot the pair $(\bar{H}_0,\text{M})$ as a blue dot in the plots displayed in the Figure~\ref{fig:robustness_test}. 
Our results show that the mean value of $H_0$ remains consistent across the different subsets, averaging close to the value obtained in our main analysis, that is, $H_0 \simeq 73$ km/s/Mpc. 
The distribution of $H_0$ values from $1000$ Monte Carlo simulations, for each subset size, predominantly falls between $72$ and $74$, in units km/s/Mpc, as shown in Figure~\ref{fig:robustness_test}, with $72$ being the value obtained in hemispheres with the highest number of SNe and $74$ from hemispheres with the lowest number. 
These findings support the conclusion that the value of $H_0$ is independent of the number of SNe within the hemisphere, and that the anti-correlation found appears to be coincidental. 

Additionally, from the $1279$ supernovae, we select samples of $309$, $512$, and $700$ SNe and estimate the three cosmological parameters: $H_0$, $\Omega_m$, and $M_b$. We then repeat the calculation for different sample sizes: $1$, $300$, $400$, $500$, $600$, $700$, $800$, $900$, $1000$, and $1100$ SNe. Our results indicate that even for low numbers of SNe, the estimated parameter values remain largely independent of the number of selected SNe. 
The observed differences can be attributed to statistical noise, which depends on the sample size.

\section{Robustness test: the other \texorpdfstring{$z_{\text{min}}$}{} cases}\label{z-min-cases}

In this appendix, we investigate the large-angle signature for 
other $z_{\text{min}}$ cases, specifically supernova samples with $z_{\text{min}} = 0.01$ and $z_{\text{min}} = 0.02$. 
To this end, we examine the dipole behavior of the $H_0$- and $\Omega_m$-maps for these samples. 

\begin{figure}[htbp]
\centering
\includegraphics[width=0.78\linewidth]{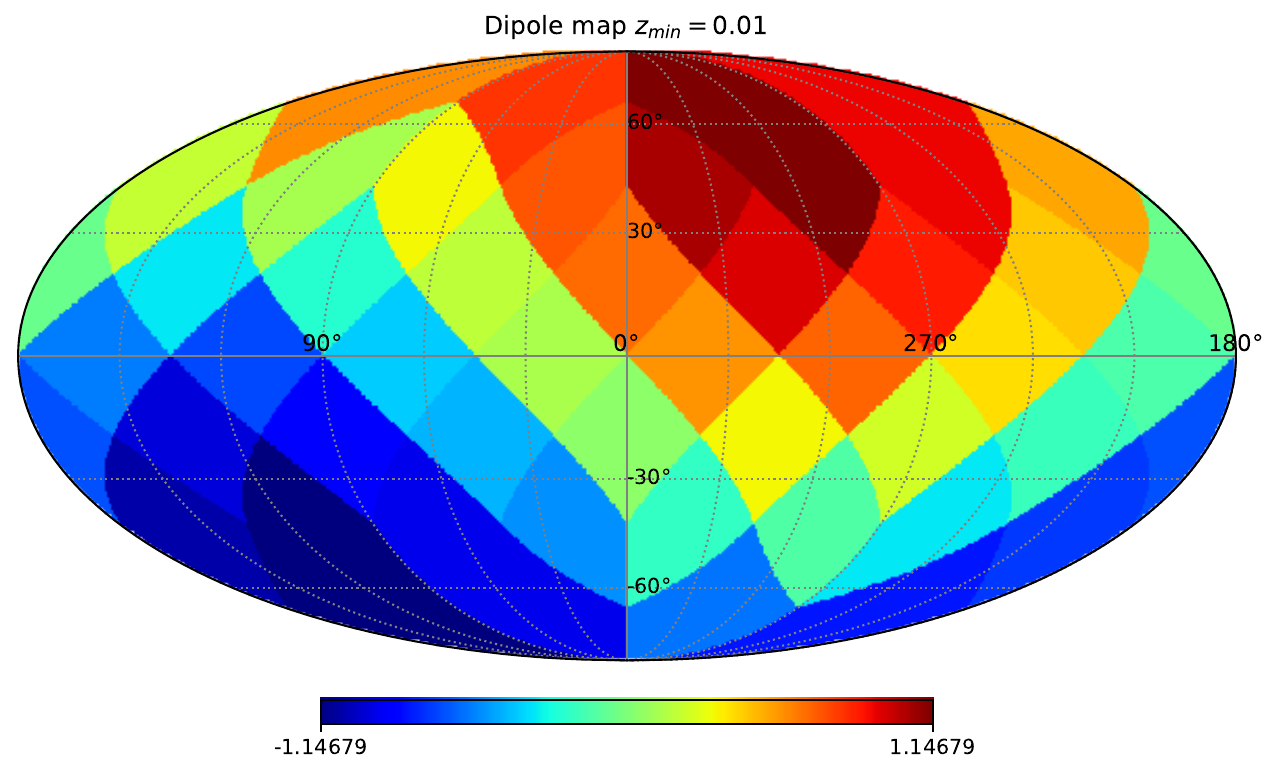}
\includegraphics[width=0.78\linewidth]{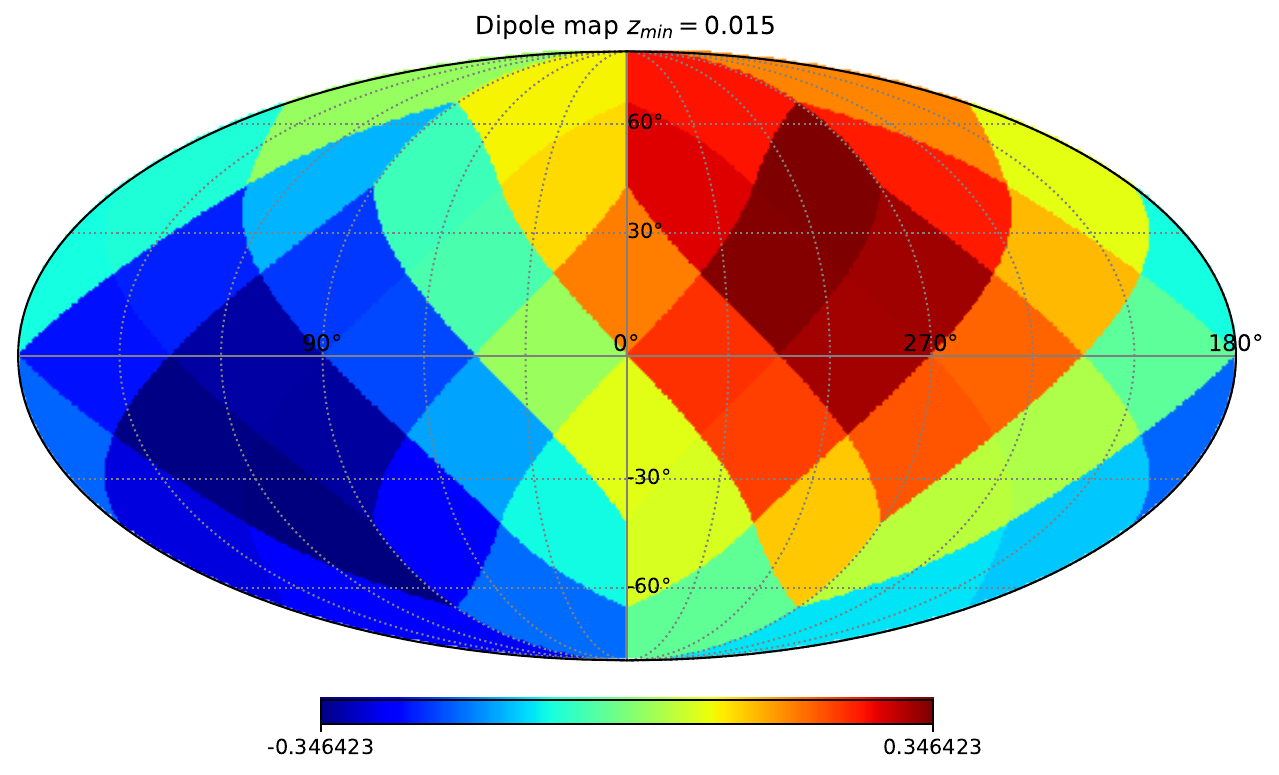}
\includegraphics[width=0.78\linewidth]{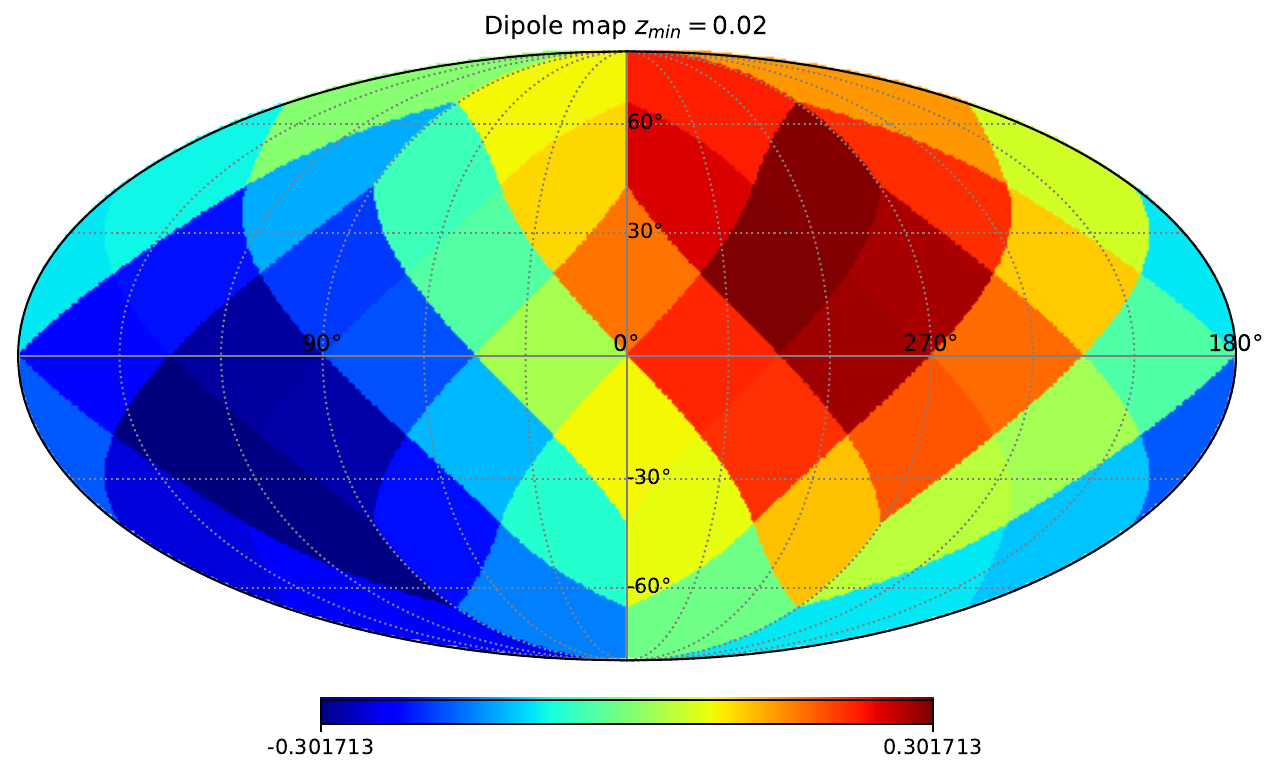}
\caption{From top to bottom, the corresponding dipole components 
of the $H_0$-maps obtained for the SNe datasets with 
$z_{\text{min}} = 0.01$, $z_{\text{min}} = 0.015$, 
and $z_{\text{min}} = 0.02$, respectively.}
\label{fig:h0_map_dipole}
\end{figure}

\begin{figure}[htbp]
\centering
\includegraphics[width=0.78\linewidth]{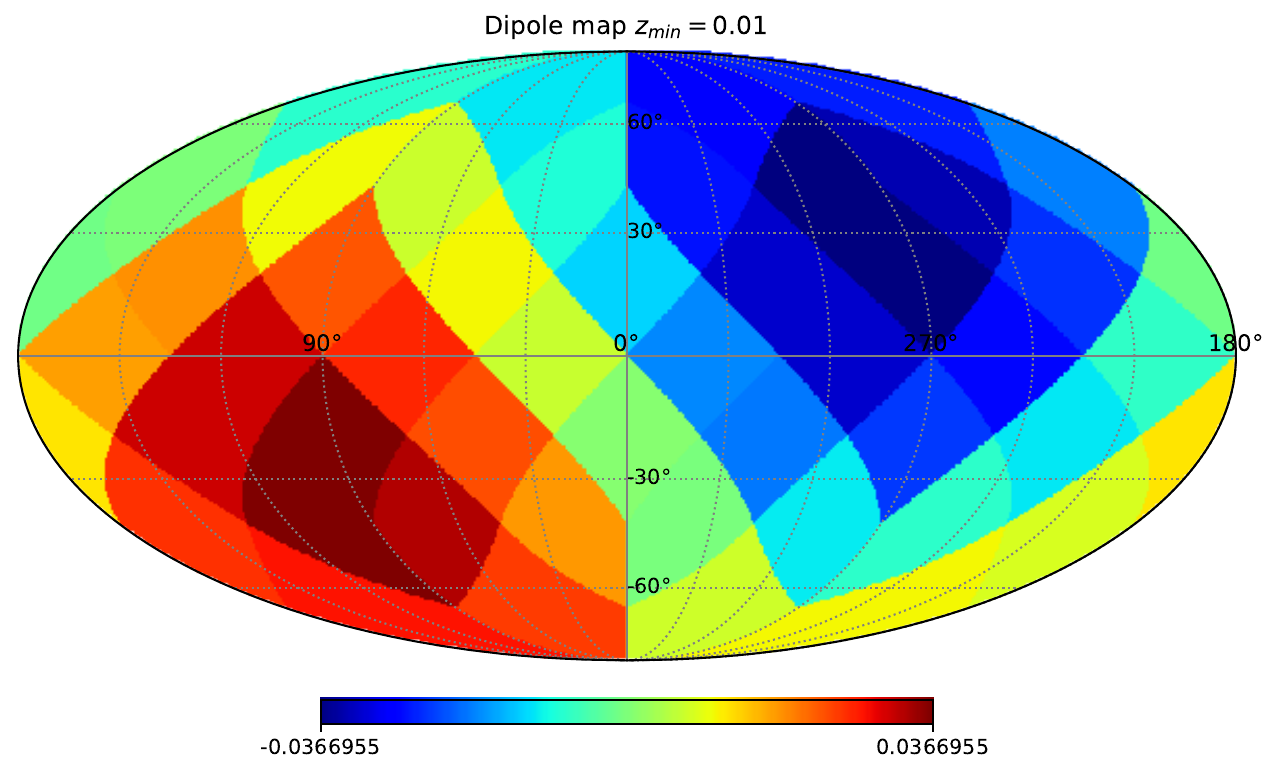}
\includegraphics[width=0.78\linewidth]{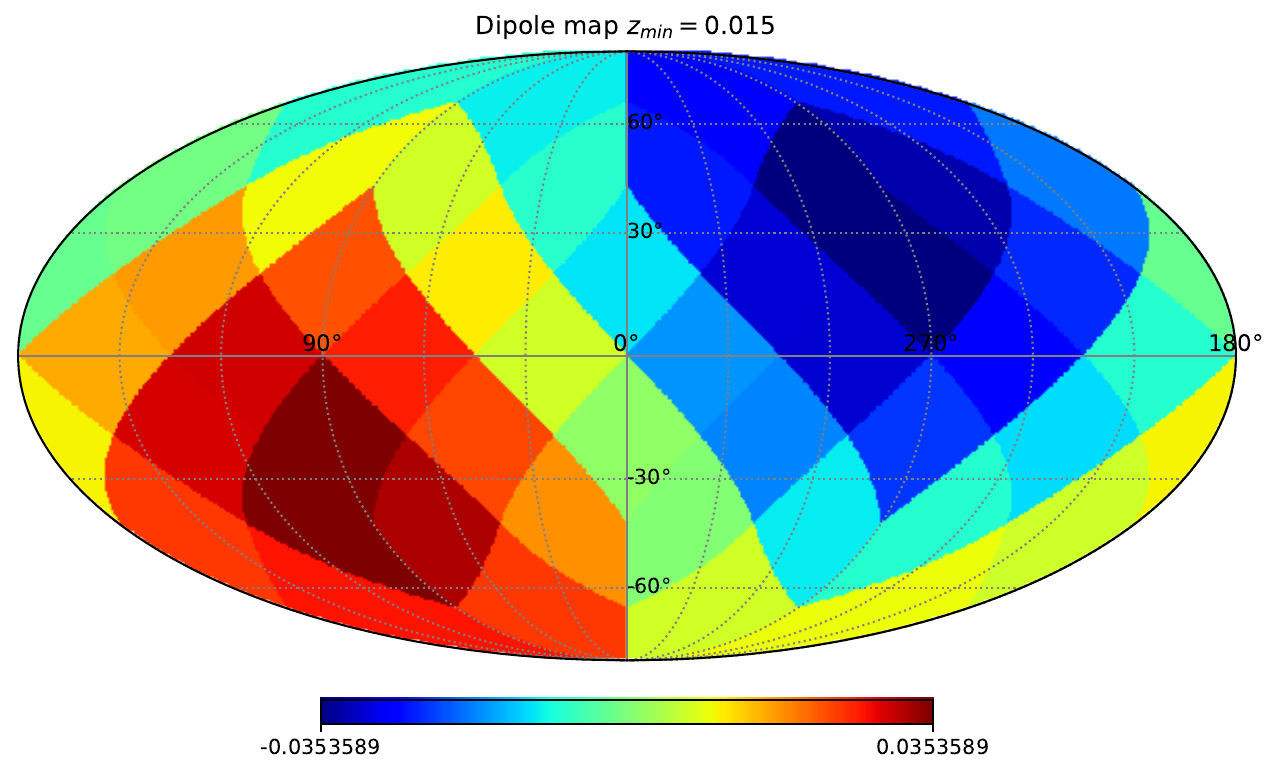}
\includegraphics[width=0.78\linewidth]{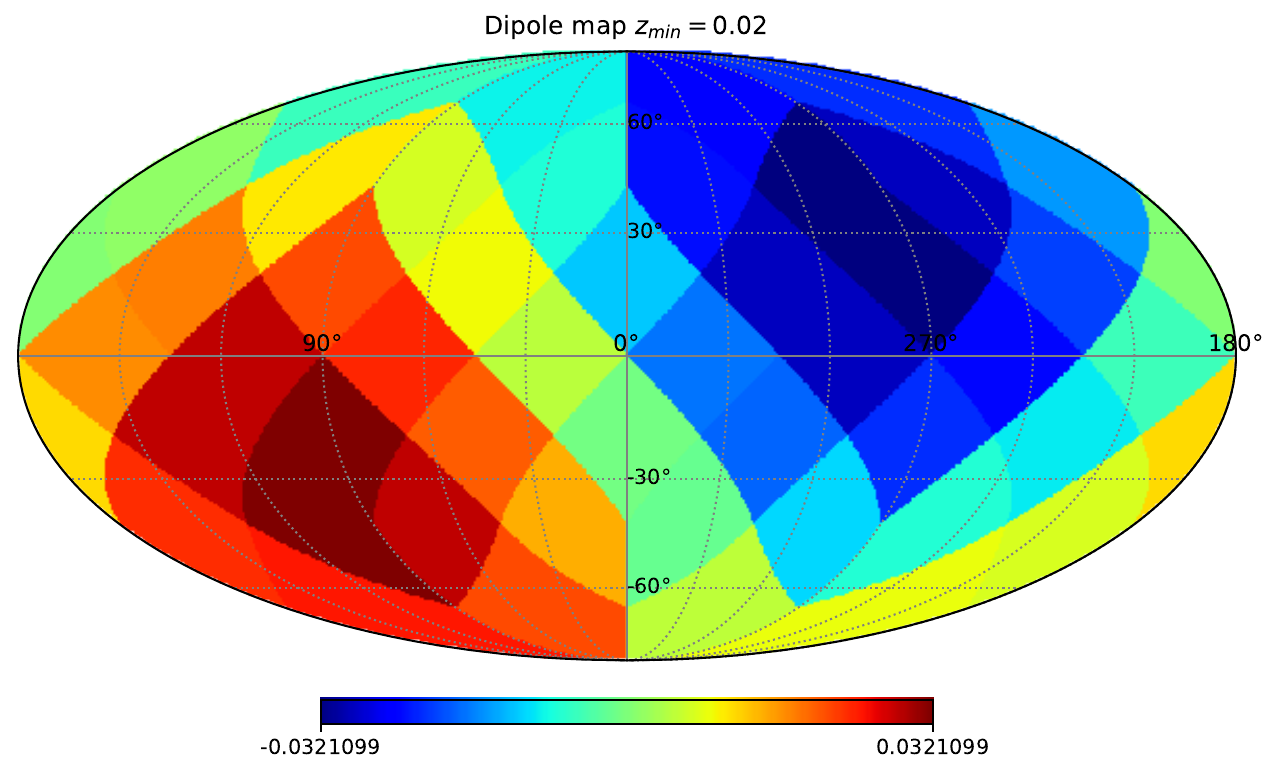}
\caption{
From top to bottom, the corresponding dipole components 
of the $\Omega_m$-maps obtained for the SNe datasets with 
$z_{\text{min}} = 0.01$, $z_{\text{min}} = 0.015$, 
and $z_{\text{min}} = 0.02$, respectively.
}
\label{fig:Om_map_dipole}
\end{figure}

\begin{figure}[htbp]
\centering
\includegraphics[width=0.78\columnwidth]{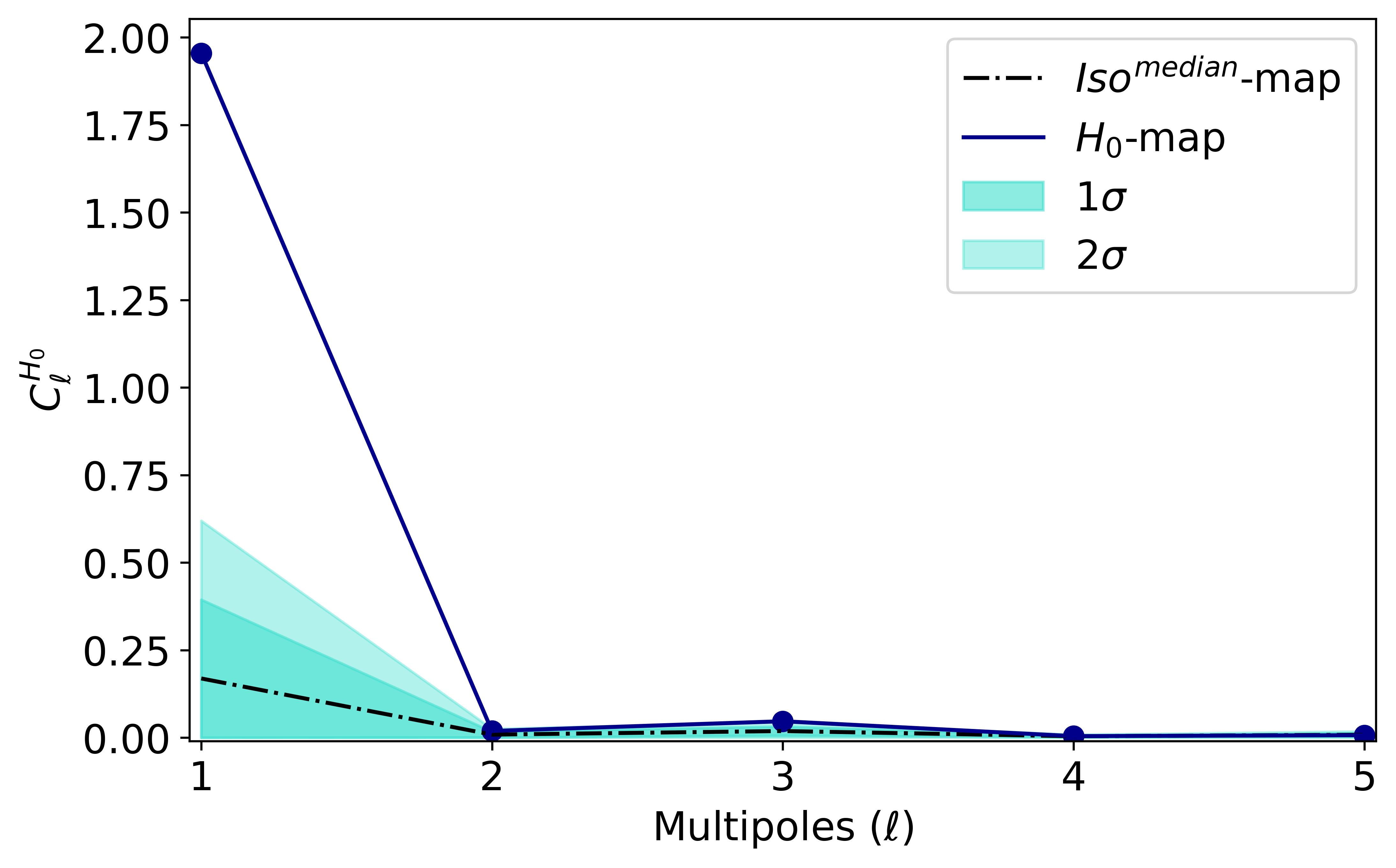}
\includegraphics[width=0.78\columnwidth]{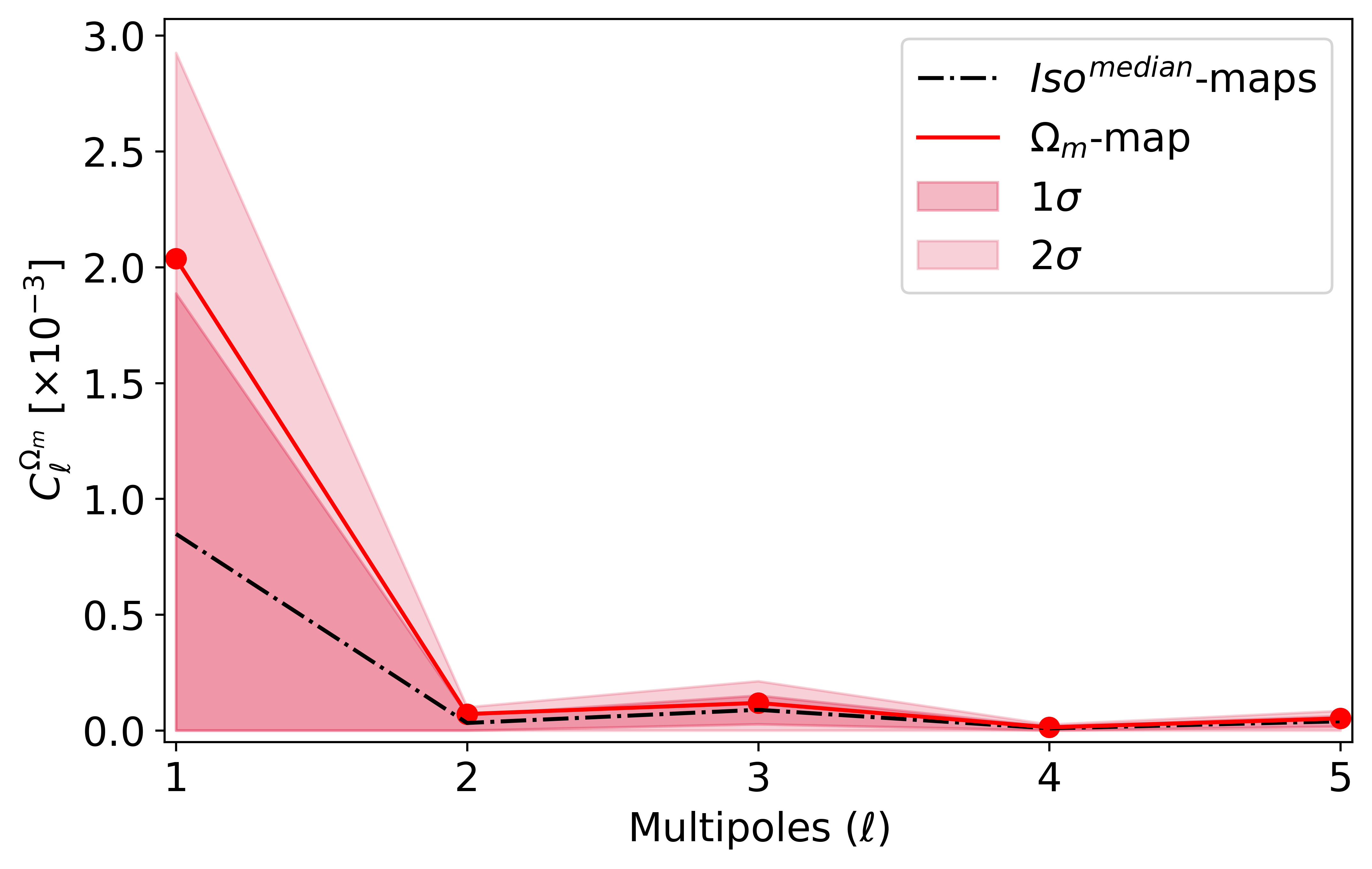}
\caption{The plots illustrate the angular power spectra of the $H_0$ and $\Omega_m$ maps alongside their corresponding ISO-maps for $z_{min}=0.01$. 
The shaded regions represent the 
1$\sigma$ and 2$\sigma$ confidence intervals obtained from the ensemble of 1000 ISO-maps. 
The comparison reveals how the observed maps deviate from statistical isotropy, providing insights into possible directional dependencies in $H_0$ and $\Omega_m$.}
\label{fig:iso_z001}
\end{figure}

\begin{figure}[htbp]
\centering
\includegraphics[width=0.78\columnwidth]{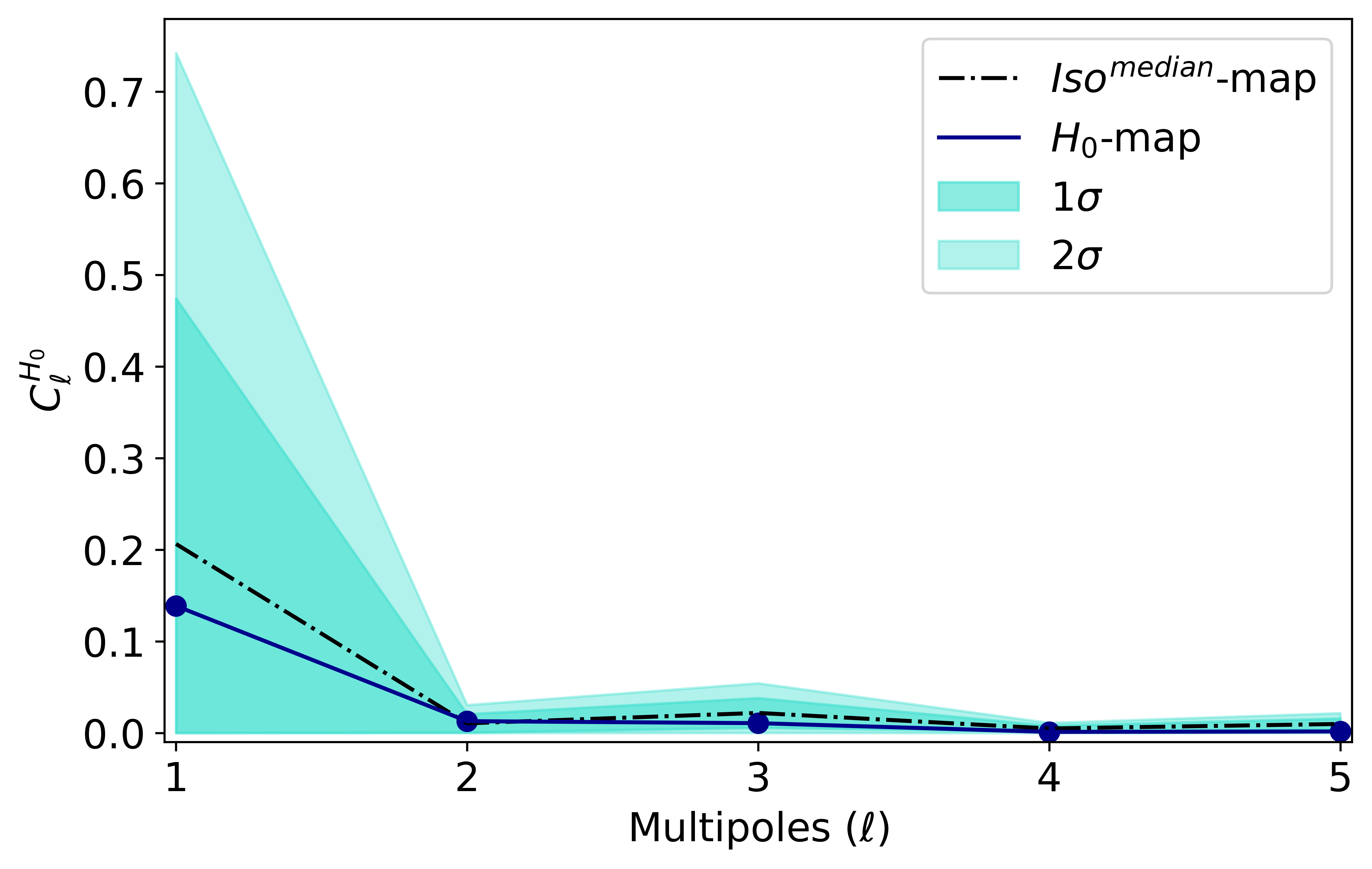}
\includegraphics[width=0.78\columnwidth]{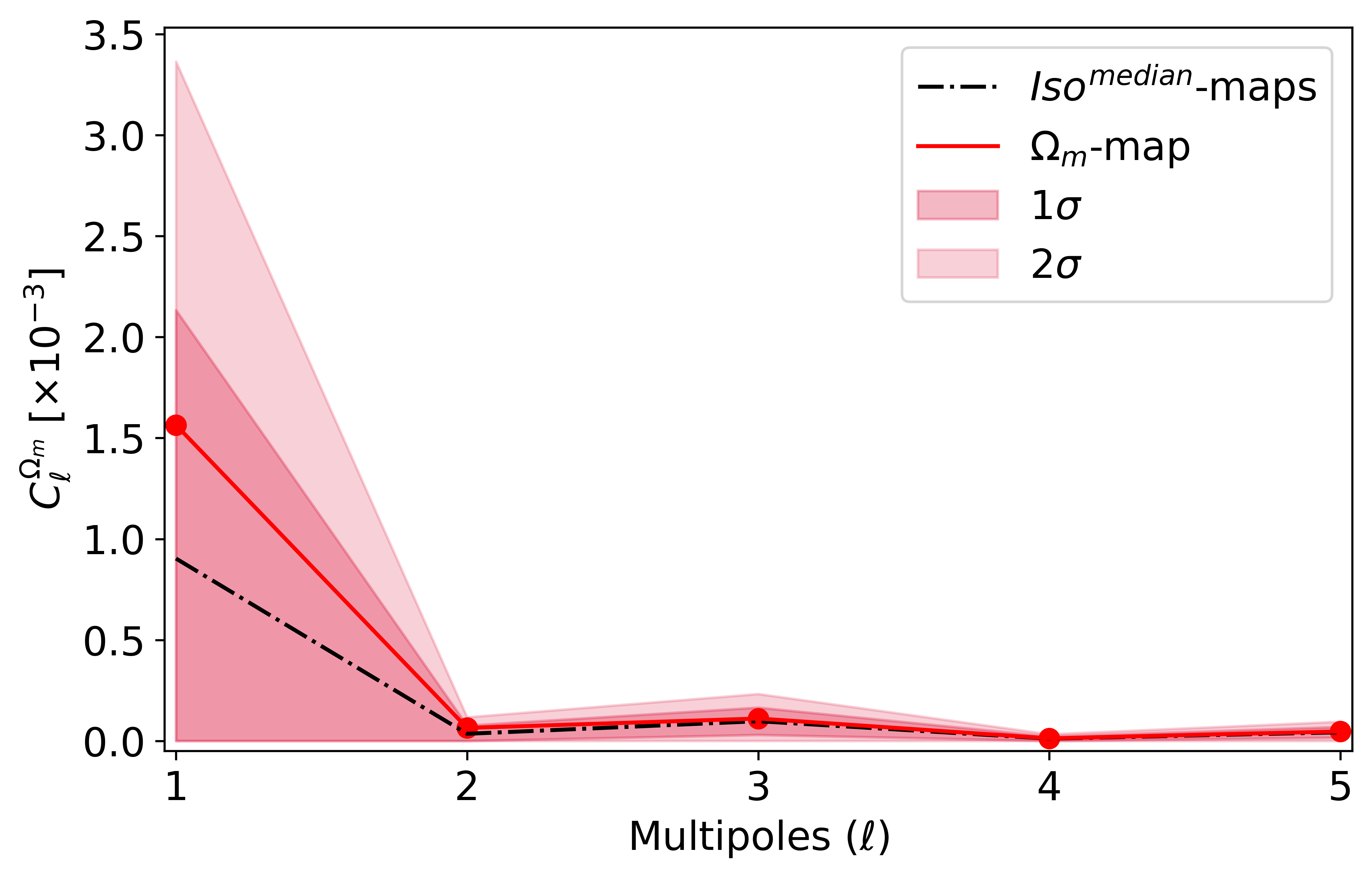}
\caption{The plots illustrate the angular power spectra of the $H_0$ and $\Omega_m$ maps alongside their corresponding ISO-maps for $z_{min}=0.02$. 
The shaded regions represent the 
$1\,\sigma$ and $2\,\sigma$ confidence intervals obtained from the ensemble of 1000 ISO-maps. 
The comparison reveals how the observed maps deviate from statistical isotropy, providing insights into possible directional dependencies in $H_0$ and $\Omega_m$.}
\label{fig:iso_z002}
\end{figure}

The dipole components of our maps results are shown in Figures~\ref{fig:h0_map_dipole} and \ref{fig:Om_map_dipole}
The statistics and dipole directions of these maps are provided in tables~\ref{tab:h0_median} and~\ref{tab:Omega_median} for the $H_0$-maps 
and the $\Omega_m$-maps, respectively. In Figures ~\ref{fig:iso_z001} and \ref{fig:iso_z002}, the angular power spectra for the parameter directional maps are shown.

The values observed in the tables and  Figures ~\ref{fig:h0_map_dipole}-\ref{fig:iso_z002}  show the effect of the low-$z$ data on the dipolar pattern of the  parameter maps, likely caused by large peculiar velocities in the local universe~\citep{Avila2023,Lopes2024,Sorrenti24,Sorrenti2024c, Swati2025,Courtois2025}.

\end{document}